\documentclass[]{aa}
\usepackage{graphicx}
\usepackage{txfonts}
\usepackage{url}
\usepackage{natbib}
\usepackage{xcolor}
\begin{document}
\title{Open questions on prominences from coordinated observations by IRIS, Hinode, SDO/AIA, THEMIS, and the Meudon/MSDP}
\author{B.Schmieder\inst{1}
\and
       H.Tian \inst{2} 
       \and
        T.Kucera\inst{3}
              \and
       A.L\'opez Ariste\inst{4}
        \and
        N.Mein\inst{1}
           \and
       P.Mein\inst{1}
   \and
        K.Dalmasse\inst{1}
\and
        L.Golub \inst{2}
        }
\institute
{Observatoire de Paris, Section de Meudon, LESIA, France\\
\email{brigitte.schmieder@obspm.fr}
\and
Harvard-Smithsonian Center for Astrophysics, Cambridge, MA, USA
\and
NASA, GSFC, MD, USA
\and
THEMIS, CNRS, LaLaguna, Tenerife, Spain
}

\date{Received date / Accepted date }
\titlerunning{Prominence in Mg II lines}
\authorrunning{Schmieder et al.}

\abstract
{ A  large prominence  was  observed 
on September
24, 2013, for three hours (12:12 UT -15:12 UT) with 
the  newly launched (June 2013) Interface Region Imaging Spectrograph (IRIS), THEMIS (Tenerife),  the Hinode Solar Optical Telescope (SOT), the Solar Dynamic Observatory's Atmospheric Imaging Assembly (SDO/AIA), and the Multichannel Subtractive Double Pass spectrograph (MSDP)  in the  Meudon Solar Tower.
}
{The aim of this work is to study the dynamics of the prominence  fine structures in multiple wavelengths to understand their formation.} 
{ The spectrographs IRIS and MSDP  provided  line profiles with a high cadence  in Mg II and in H$\alpha$ lines.
}
{The  magnetic field is  found to be globally horizontal with a relatively weak  field strength (8-15 Gauss).  
The Ca II movie reveals turbulent-like motion that is not organized in specific parts of the prominence. 
 On the other hand,  the Mg II line profiles  show multiple peaks well separated in wavelength. 
 Each peak corresponds to a Gaussian profile,  and not  to  a reversed profile as  was expected by the  present non-LTE radiative transfer modeling. 
}
{Turbulent fields on top of the macroscopic horizontal component of the magnetic field supporting the prominence give rise to the complex dynamics of the plasma. 
  The plasma with  the high velocities (70 km/s to 100 km/s  if we take into account  the transverse velocities). 
 may correspond to condensation of plasma along  more or less  horizontal threads of the  arch-shape structure visible  in  304 \AA.
  The steady flows (5  km/s)  would correspond  to a more quiescent plasma (cool and prominence-corona transition region) of the prominence   
 packed into dips in  horizontal magnetic    field lines.  The very weak secondary peaks in the Mg II profiles may reflect the turbulent  nature of parts of the 
 prominence.}


 \maketitle


\section{Introduction}
Prominences, filaments when observed on the disk, are large structures in the solar corona filled with cool dense plasma  suspended above 
 magnetic polarity inversion lines \citep[see reviews of][] {MacKay2010,Labrosse2010,IAUS3002014}. 
The formation, structure, and evolution of solar filaments and prominences are  an important part of our understanding of coronal physics. 
 Recent observations show that surface motions acting on     magnetic fields  which  are non potential may play an important role in the formation of large scale filaments \citep{vanBallegooijen1989}, while flux rope emergence may be part of the formation of small filaments \citep{Okamoto2008}.
Despite daily observations of filaments and prominences with coronagraphs and the Solar Dynamic Observation (SDO) their formation process is still unclear. 

Many questions about the formation of filaments are still debated.  Is  filament formation  due to condensation of coronal material  along flux tubes in coronal cavities   \citep{Karpen2005,Luna2012}? Does the flux rope corresponding to a filament  lift up through the photosphere by levitation process \citep{Okamoto2008}  or is it formed by successive reconnections between magnetic field lines \citep{vanBallegooijen1989,Schmieder2006}?
Even with the new SDO observations the structure of the prominence-corona transition region  (PCTR) is still unclear \citep{Parenti2012}.   The observations performed so far   have not resolved these questions.

Prominences observed in different lines or  pass band filters look different  depending not only on the formation temperature of the line, but also on the optical thickness of the line. These differences can lead to confusion, but can also be important tools to enhance our understanding of prominences.

New instrumentation (Hinode/SOT, SDO)  has  revealed new details concerning the highly dynamic and complexly structured nature of prominences. Considering these data, \citet{Priest2014} concluded that two magneto-hydrodynamic  systems may  be considered to explain prominences: one of them for the global magnetic scale and the second  one relevant to turbulence. 

With multiwavelength observations  it is possible to study the dynamics of  the structures in a wide range of temperatures. \citet{Berger2012} proposed that the bubbles and rising plumes observed with SOT are hot and fill the cavity with plasma before condensation. However, there is a controversy  about the existence of thermal  instabilities explaining the rising structures in prominences. The bubbles may be due to a separatrix around emerging flux and the  dark  rising structures  just  open windows through the prominence, allowing us to see the background corona \citep{Dudik2012,Gunar2014}.  Spectroscopic data obtained by SOHO/SUMER  and Hinode/EIS confirmed  that the bubbles did not contain hotter plasma  than the well expected prominence-corona transition region (PCTR) \citep{Berlicki2011,Labrosse2011,Gunar2014}.  Nevertheless, the low cadence and low resolution of the data may be a reason why such results were obtained.

There is also controversy concerning the orientation of the magnetic field in the structures observed in  prominences. The statistical measurements of the magnetic field in a large number of prominences obtained in the past by \citet{Leroy1984,Bommier1994}, and the recent measurements  obtained  for a Hinode/SOT prominence with THEMIS \citep{Schmieder2013} indicate that the magnetic fields are horizontal. However, the SDO  and Hinode images still give the impression that the plasma structures are mainly vertical. Measuring the velocity vectors in such structures is a good way to estimate  their inclination versus the vertical, as was done recently by \citet{Schmieder2010} using H$\alpha$ line observations in a hedge-row prominence. The transverse velocities were found to be on the same order as  the Doppler shifts indicating an inclination of 45 degrees  for the structures towards the vertical.


The Interface Region Imaging Spectrograph (IRIS) with  its high spatial resolution (pixel size  0.167 arcsec,  resolution=0.4 arcsec)  and  its incredible high spectral resolution  (0.05 \AA) was launched in June 2013 
\citep{DePontieu2014}  and  is a very suitable instrument  with which  the dynamics of  the fine structures of prominences can be studied. We had the opportunity to observe a large prominence on September 24, 2013, during the first 60 days of science observations of IRIS with  Hinode/SOT, the THEMIS vector magnetograph in Tenerife,  and the MSDP in the Meudon solar tower. 
This campaign  provided an excellent opportunity to put the IRIS data in a global perspective associated with  other instrument results.

In the next section, we present the characteristics of the instruments that we used. Section \ref{s:results} presents the results concerning the magnetic field and the velocity vectors in different lines. In the last section we discuss the results and conclusions, including the importance of having spectroscopic and spectro-polarimetry diagnostics for determining  the true velocity of the plasma in the fine structures. 

%
 \begin{figure} 
 \centerline{\includegraphics[width=0.4\textwidth,clip=]{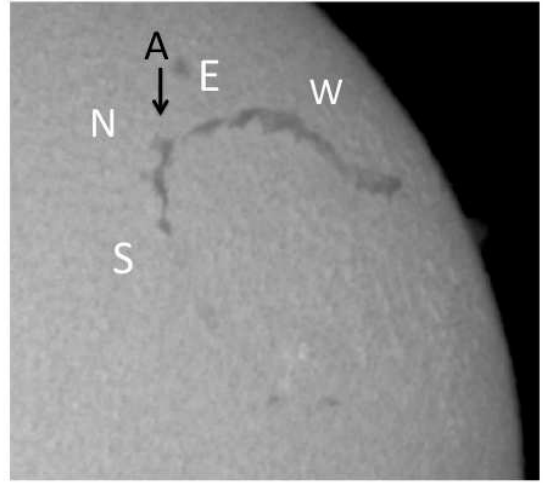}}
 \caption{ Filament observed on September 20, 2013,  with two sections oriented east-west (EW)  and north-south (NS) respectively (Meudon spectroheliograph survey). The letter A indicates the junction between the two sections and is  clearly a foot of the filament.}
 \label{meudon}
 \end{figure}

%
 \begin{figure*} 
  \centerline{\includegraphics[width=0.50\textwidth,clip=]{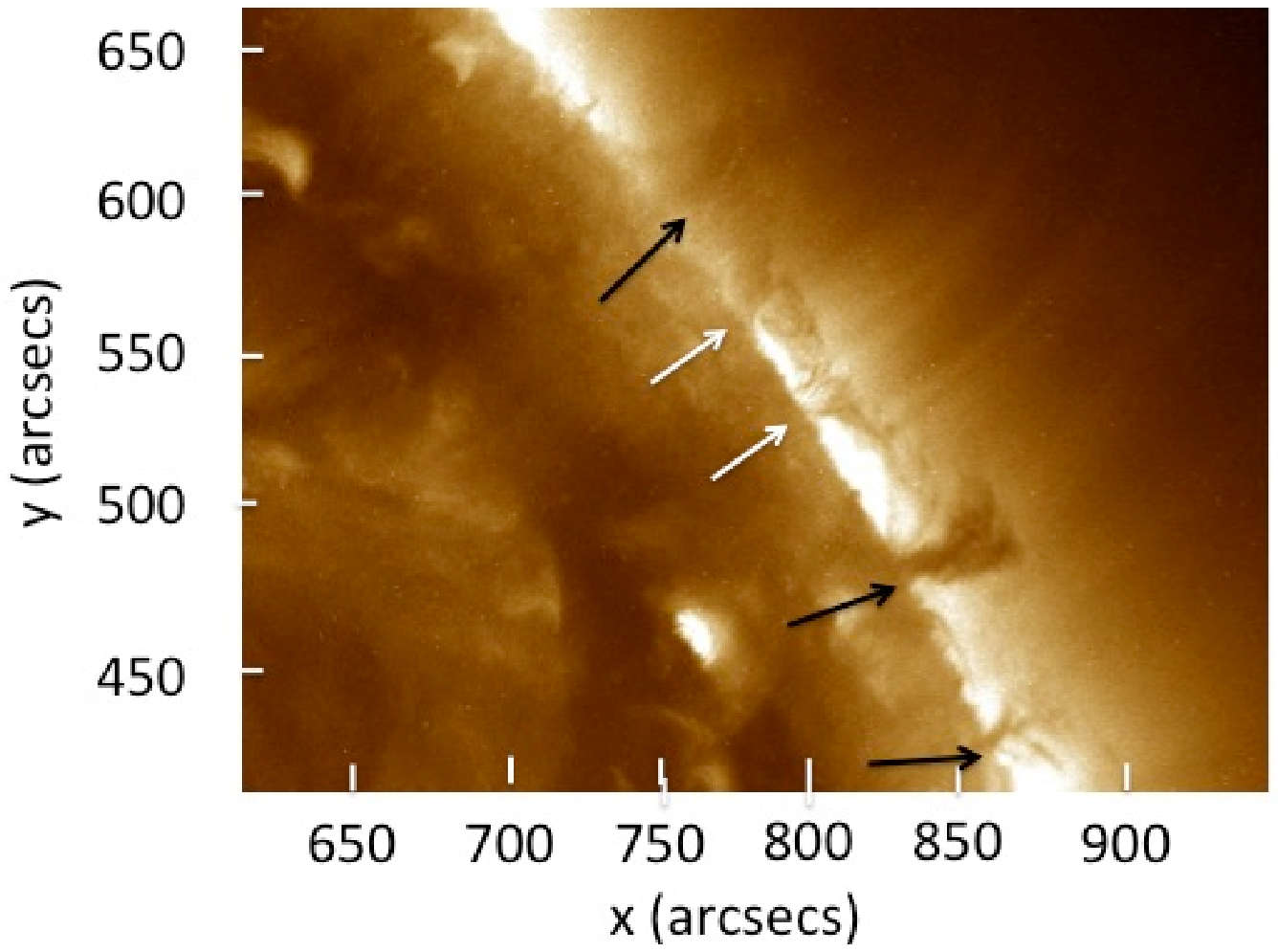}
 \includegraphics[width=0.5\textwidth,clip=]{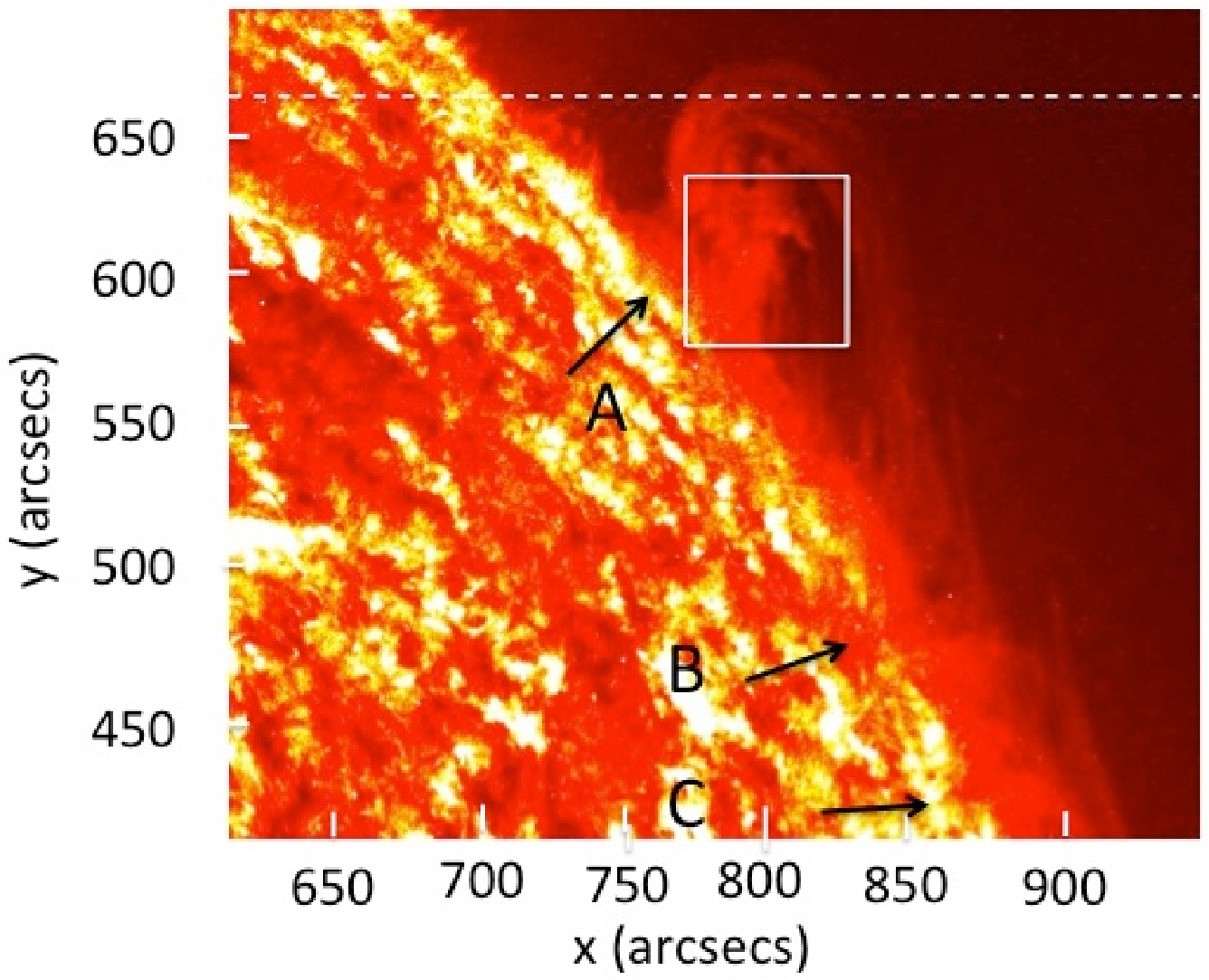}}
   \centerline{\includegraphics[width=0.50\textwidth,clip=]{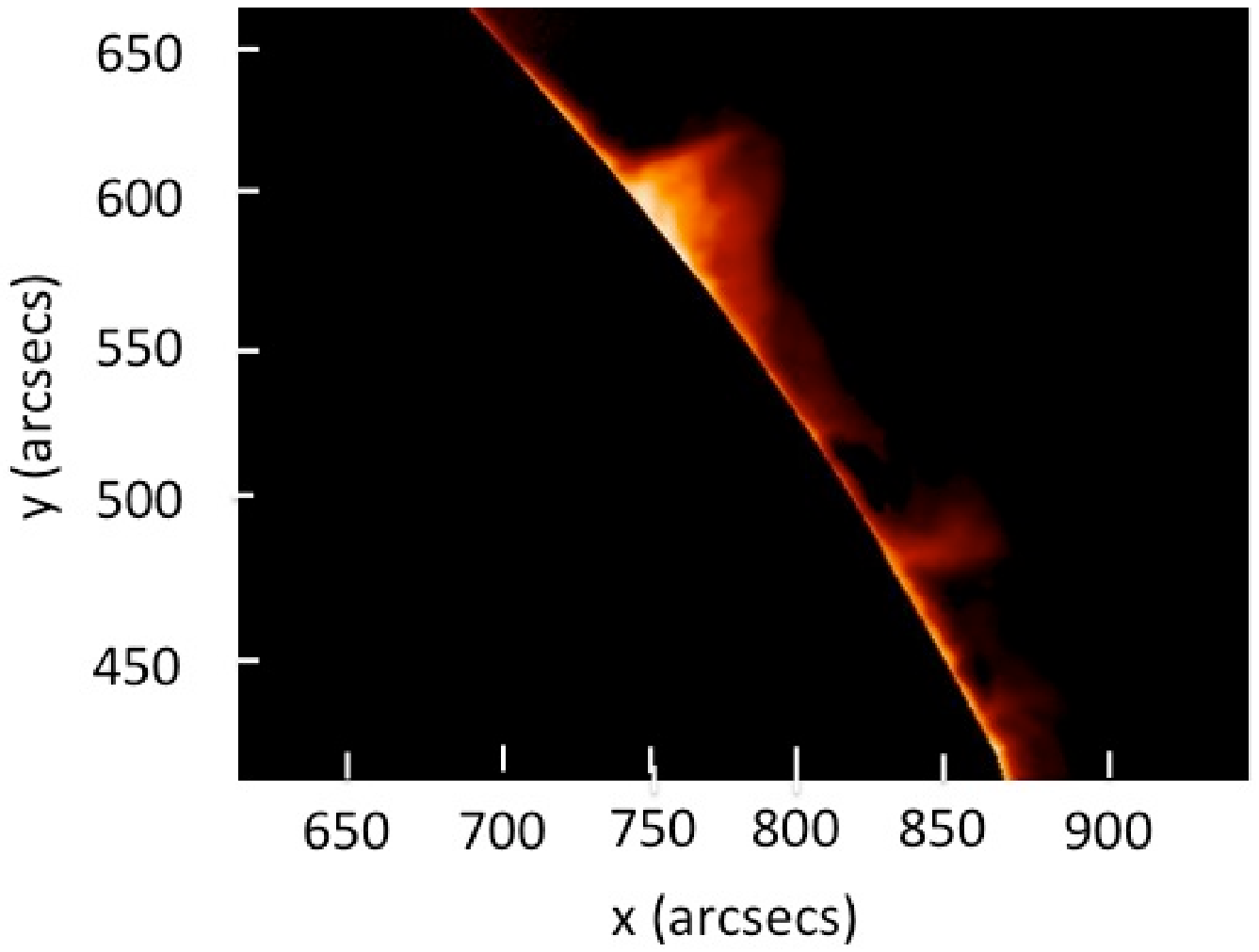}
    \includegraphics[width=0.5 \textwidth,clip=]{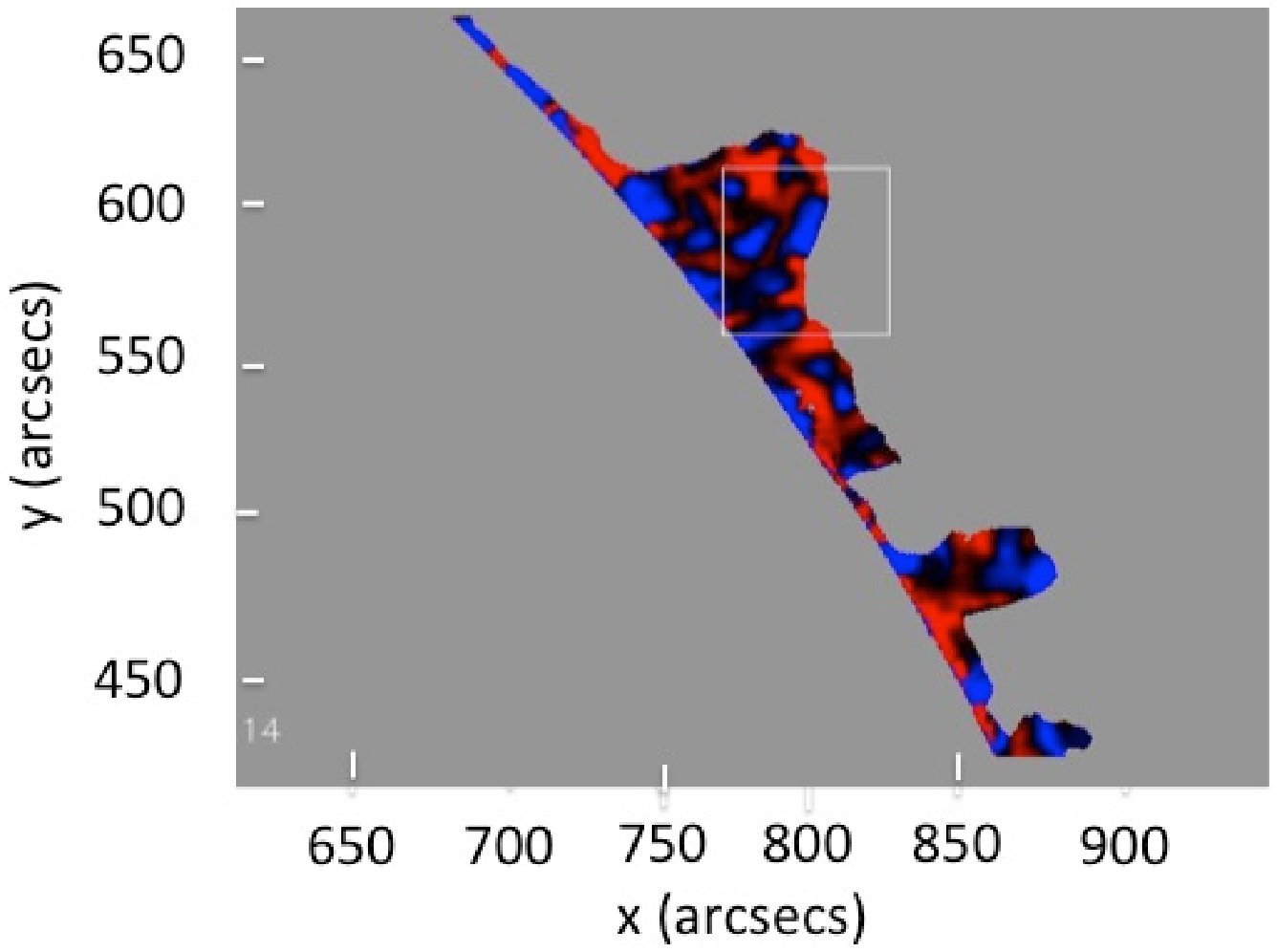}} 
   \caption{({\it top  panels}) Prominence observed on September 24,  2013, at 12:22 UT in  193 \AA\  ({\it left}), and in 304 \AA\   ({\it right}) from SDO/AIA.  The black arrows and the letters A, B, C indicate  feet of the 304 \AA\ prominence. The white arrows are other feet only present in 193 \AA.   ({\it bottom   panels}) The prominence observed in H$\alpha$ by  the MSDP  ({\it left}) intensity,  and  ({\it right})  Doppler shift  in H$\alpha$ $\pm$ 0.3 \AA\,Ê black/white (blue/red on the online version) indicate blueshift/redshift limited to $\pm$ 2 km/s).  The boxes in the 304 \AA\ image and in the Doppler shift image  are  approximately  the field of view of the IRIS slit jaws (50 x  50 arcsec).}
 \label{msdp2}
 \end{figure*}

\begin{figure*} 
\centerline{\includegraphics[width=0.8\textwidth,clip=]{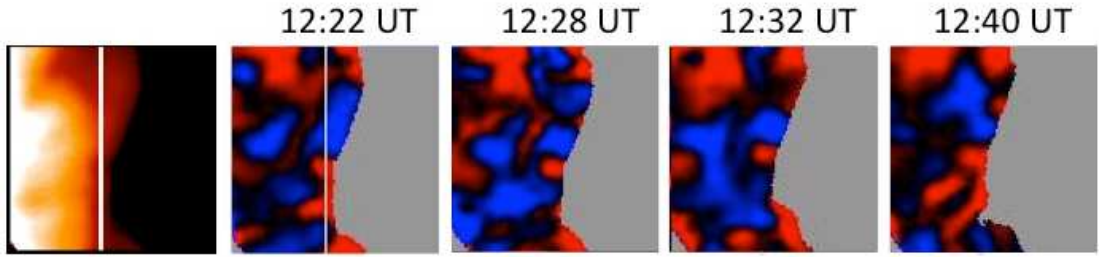}}
 \caption{H$\alpha$ intensity image (left) and four  Doppler shift images  obtained in  H$\alpha$ $\pm$ 0.3 \AA\,Ê (MSDP) (right) in the FOV of IRIS. The  Mg II h spectra at the same times are presented in the left panels of Figure \ref{iris1}.
 Black/white (blue/red on the online version) are the blueshifts/redshifts limited to $\pm$ 2 km/s. The white vertical  line approximately represents  the location of the  IRIS slit 1  used  in  Figure \ref{cuts}.}
 \label{msdp3}
 \end{figure*}

\section{Observations}
\subsection{Description}
 
 \begin{figure*} 
 \centerline{
 \includegraphics[width=0.42\textwidth,clip=]{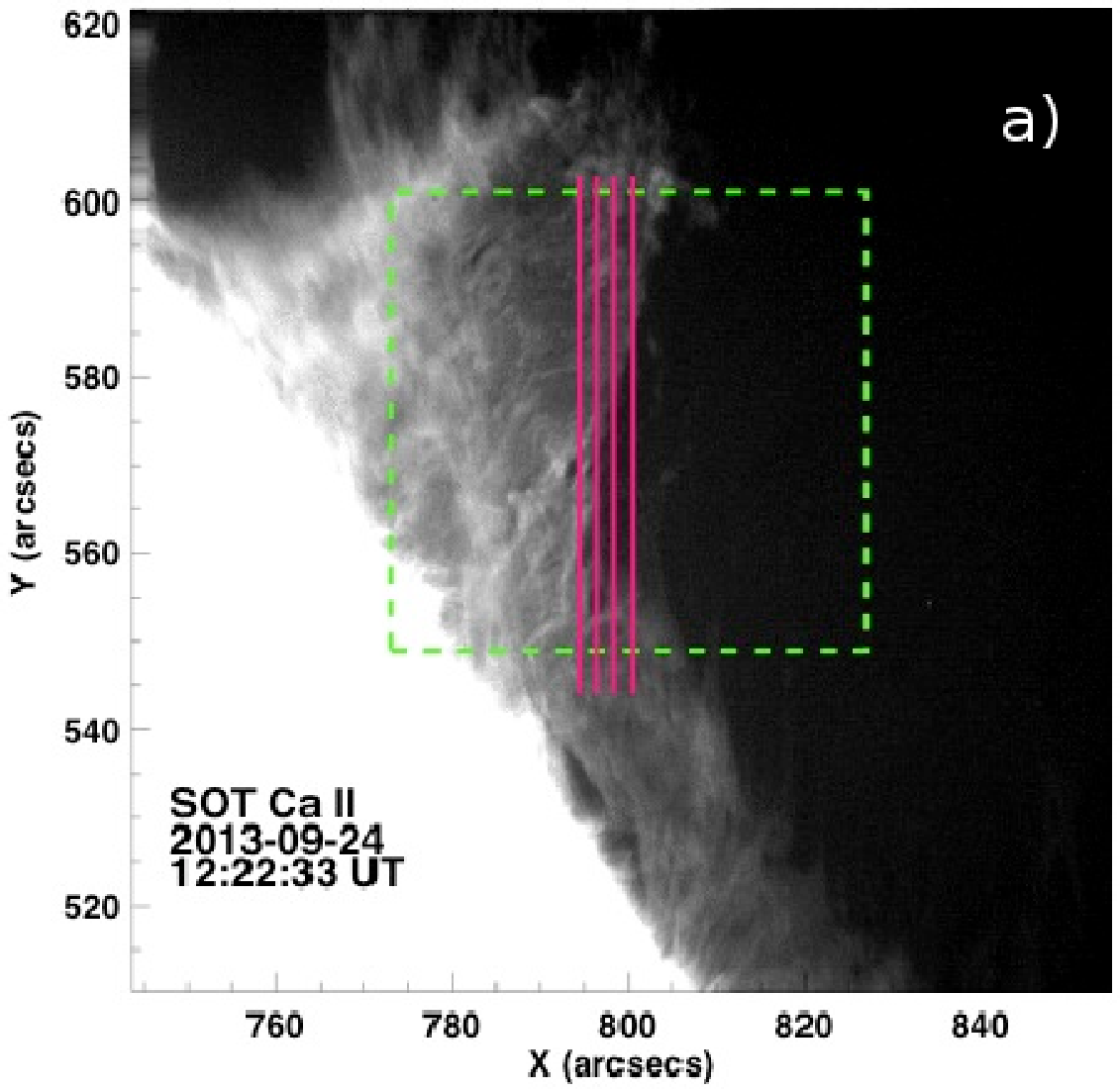}
   \includegraphics[width=0.43\textwidth,clip=]{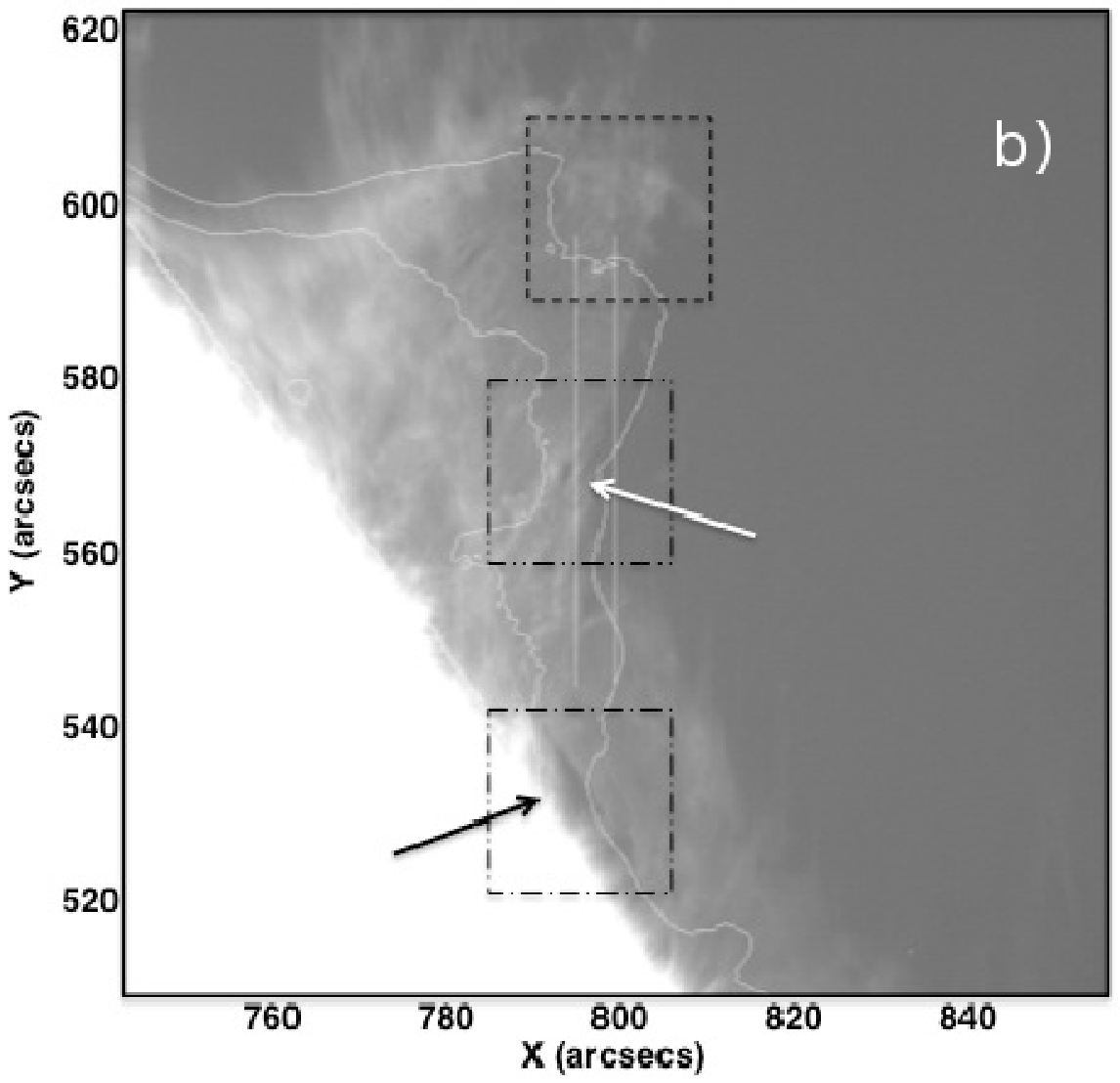}}
  \centerline{
   \includegraphics[width=0.8\textwidth,clip=]{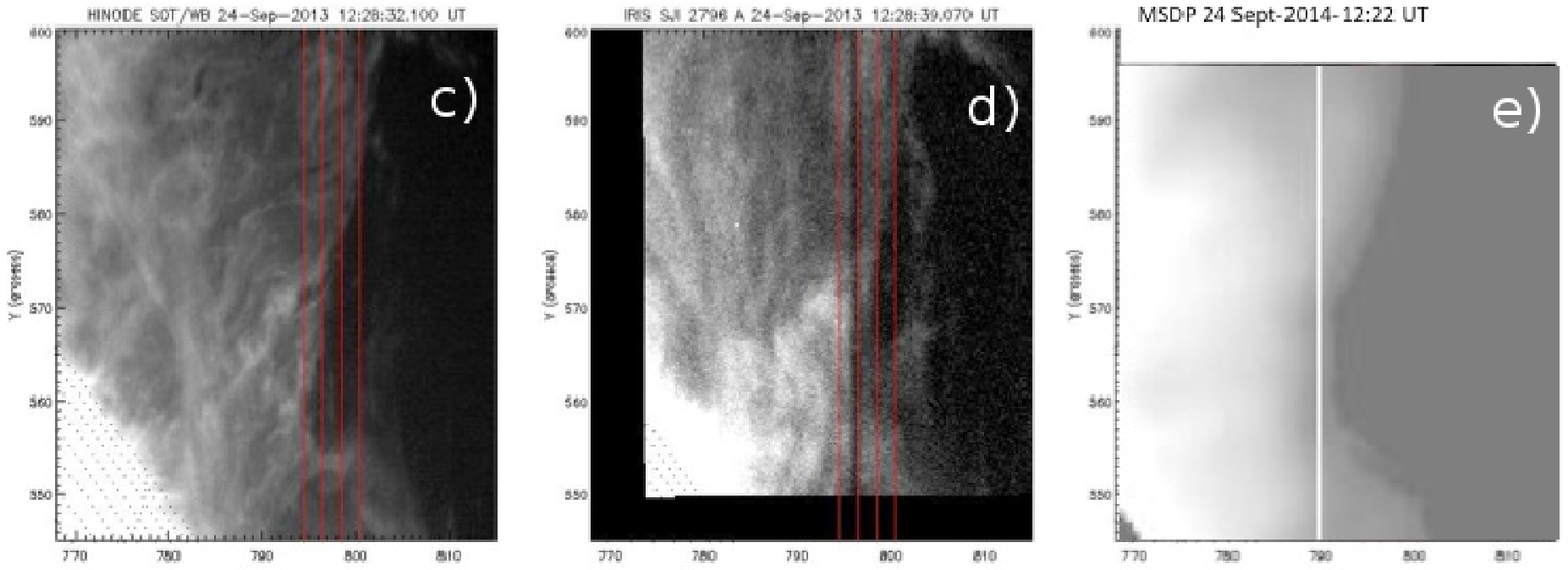}}
  \caption{(a)  Hinode/SOT Ca II  at 12:22 UT  with a box representing  the field of view of IRIS slit jaw and  four vertical red  lines, the four IRIS slits, and (b) the same  Ca II image overlaid with the contours of the H$\alpha$ intensity of MSDP and three   small boxes: the top box indicates  a turbulent region, the middle  box  with the white arrow a small dark bubble, the bottom box with the black arrow
  a large limb bubble.  An online  movie of Hinode/SOT  Ca II shows the temporal evolution of  the image  (b); the boxes drawn in (b) help to follow the different dynamics observed in  the prominence.   ({\it bottom panels}) Images obtained with  three different instruments in the IRIS field of view ( c) in Ca II (SOT), (d) in Mg II (IRIS slit jaw),  and (e) in H$\alpha$ (MSDP).  }
 \label{iris_MSDP}
 \end{figure*}

A long filament with  two perpendicular sections (section   EW  N35$^\circ$-38$^\circ$ and the section between N15$^\circ$ and N30$^\circ$ ) was crossing the limb on September 24, 2013 (Figure \ref{meudon}).  The intersection (A) between these two sections was just at the limb on September 24 and corresponds to a  filament foot  or barb (see the arrow).

The prominence as observed in 304 \AA\ with SDO/AIA 
consists of a large triangle in the north  (x=800 arcsec, y=600 arcsec),  which is probably part of the EW section  integrated along its axis, and  the feet  A, integrated along the line of sight, and a bright  arch of  long  horizontal   threads parallel to the  limb (Figure \ref{msdp2}). This long arch is part of the NS section, between pillars or feet (A, B, C,  in Figure \ref{msdp2}), according to the heliographic coordinates of the filament. Material is continuously flowing along the long threads in both directions.
 In the 193 \AA\ filter the prominence appears  dark  due to the absorption of the coronal line by resonance  continua of hydrogen  and helium  at this wavelength. The dark structures in 193 \AA\  are  similar to  those in H$\alpha$ because they have similar  optical thickness \citep{Anzer2005, Schmieder2004}.  The shape of the dark absorption is very different from the emission prominence observed in 304 \AA\ (Figure \ref{msdp2}). 
The emission in 304 \AA\ is due to  the scattering of  chromospheric line radiation and partially due to the presence of  a  prominence-corona transition region
 \citep{Labrosse2010}.
We want to highlight  that the H$\alpha$ prominence (here observed in absorption in the 193 \AA\ coronal line)   has more anchorage points (feet) with the photosphere than the 304 \AA\ prominence. It confirmed that lateral extensions in EUV filaments (channels) do not  always coincide with H$\alpha$ filament feet or barbs \citep{Aulanier2002,Schmieder2014}. 

\subsection{MSDP}
The  Meudon MSDP observations  of prominences in H$\alpha$ consist of series of five  spectral images 465  arcsec x 60 arcsec with 6 arcsec  overlaps in each time sequence.  The exposure time is 100ms. Sequences of observations are done during 15 minutes with   a 30 sec cadence.  They have been  processed with the MSDP software. On September  24, ten   sequences of observations were done  between  12:09UT and 15:09 UT. We focus our study on the sequence starting at 12:22 UT,  and 12:38 UT  because it is in the interval of the IRIS observations.
In each solar point  an H$\alpha$ line profile  is obtained over a wavelength range of +/-  0.7  \AA. Doppler shifts can be computed for any wavelength in this wavelength range  with an accuracy of 0.5 km/s (Figures \ref{msdp2} and \ref{msdp3}).

\subsection{Hinode}

The Hinode \citep{kosugi07}  SOT  \citep{tsuneta08,suematsu08} consists of  a 50 cm diffraction-limited Gregorian telescope and a Focal Plane Package including the narrowband filtergraph (NFI), the broadband filtergraph (BFI), the Stokes Spectro-Polarimeter, and Correlation Tracker (CT). For this study, images were taken with a 30 sec cadence in the \ion{Ca}II H line at 3968.5~\AA\ using the BFI. The \ion{Ca} II images have a pixel size of 0.109\arcsec, with a field of view of 112$\times$ 112\arcsec\ (Figure \ref{iris_MSDP}).  

\subsection{THEMIS}
The THEMIS/MTR instrument \citep{LARS00} was used to do spectropolarimetry of the He D$_3$ line in the observed prominence. The spectrograph slit was
oriented parallel to the local limb. This direction  subsequently defined the sign of the linear polarization: positive Stokes Q means parallel to the slit and,
in consequence, parallel to the local limb. 
The observations were obtained with same setup previously described by \citet{Schmieder2013}. The use of a grid mask  with segments 15.5 arcsec wide along the slit is required to perform  accurate measurements. 
To speed the record of the full Stokes parameters, the overlap of the segments is minimized. The exposure time is two seconds.
 Rasters are obtained with steps of 2\arcsec\ from the limb to the top of the prominence.  In our case a  field of view, approximately 120\arcsec\  by 25\arcsec\ , was covered  in approximately  two hours.
The intensity map of the prominence in He D$_3$ corresponds   reliably  to the  304 \AA\ image of SDO/AIA (Figure \ref{themis1}).

\subsection{IRIS}

IRIS performed a four-step coarse raster observation from 12:14 UT to 15:13 UT on September 24, 2013. The pointing of the telescope is 798 arc sec, 573 arc sec. The spatial pixel size is 0.167 arc sec. The raster cadence of the spectral observation in
both the near ultraviolet  (NUV  2783 - 2834 \AA) and  the far ultraviolet (FUV  1332-1348  \AA\ and 1390-1406 \AA)
wavelength bands was 35 seconds. Exposure times were 8 seconds. Slit-jaw images (SJI) in the broadband  filters   
(2796  \AA\ and 1400 \AA\ )  were taken at a cadence of 18 seconds.  The 1400~\AA\ slit jaw is an integration of the FUV emission within a range of about 40~\AA\ (including the total emission of two Si IV lines). The FOV was 6x50 arc sec$^2$  for the raster and 50x50 arc sec$^2$  for the SJI. The calibrated level 2 data was used in our study. Dark current subtraction,  flat field correction, and geometrical correction have been taken into account in the level 2 data \citep{DePontieu2014}.

We mainly used the Mg II k 2796.4~\AA\ and Mg II h 2803.5~\AA\ lines and the SJI  1400 and 2796~\AA\ data for this study. The Mg II h  and k lines are  formed at  chromospheric  temperature (10$^4$ K). The SJI
2796 \AA\  filter samples emission mainly from the Mg II k  line, while emission in the 1400 \AA\  filters from the Si IV 1402 \AA\ and 1393~\AA\ lines formed in the prominence transition region (PCTR). The UV continuum  at 1400 \AA\  formed  in the lower chromosphere is not present for  a prominence observed at the limb  contrary to observations on the disk. The co-alignment between the different optical channels of IRIS was 
achieved by checking the position of  horizontal fiducial lines.


\section{Results}
\label{s:results}

%
\begin{figure} 
\centerline{\includegraphics[width=0.5\textwidth,clip=]{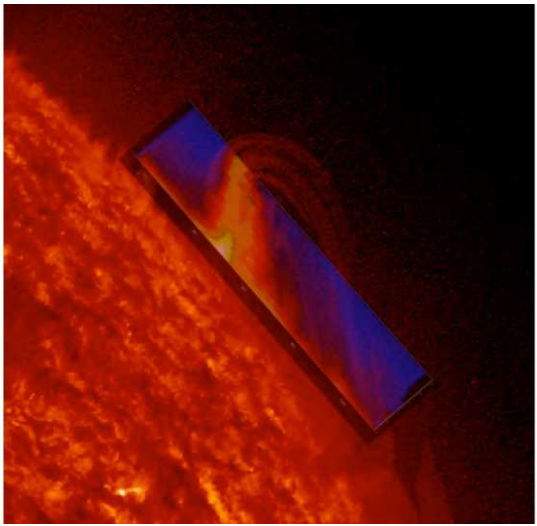}}
 \caption{THEMIS image in He D$_3$ line  overlaid on an AIA 304 \AA\, image. We note  the  large scale  304 \AA\,Êarches with bright loops.
 }
 \label{themis1}
 \end{figure}

\subsection{Co-alignment of the different observations}
The co-alignment  of the observations obtained by the different instruments  is rather difficult, particularly with IRIS. Cross-correlation between the AIA 1600 \AA\  image and the IRIS SJI images 
was used for the co-alignment between IRIS and SDO (Figure \ref{msdp2}).
  The co-alignment between SOT and IRIS is done by using the  1600~\AA\  SDO  images and the 2786 IRIS slit jaws.  The  Ca II and Mg II maps have many similarities  (Figure \ref{iris_MSDP}). The MSDP maps were aligned with SOT by using the intensity on the disk in the H$\alpha$ and Ca II lines (Figure \ref{iris_MSDP}).  The THEMIS He D$_3$ intensity map was co-aligned with AIA images in 304 \AA\  (Figure \ref{themis1}).

%
\begin{figure} 
\centerline{\includegraphics[width=0.5\textwidth,clip=]{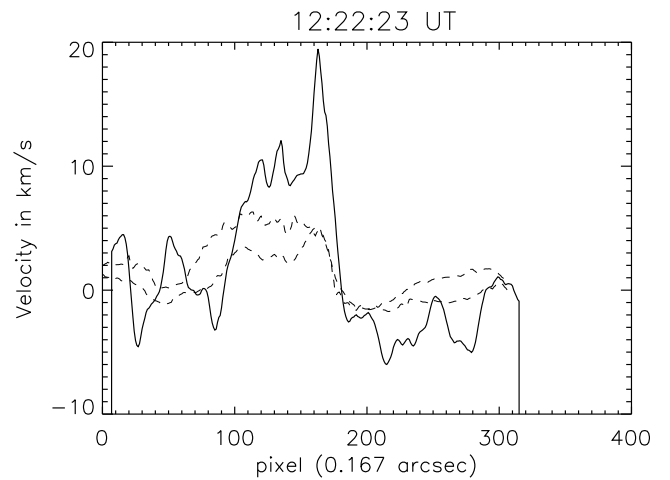}}
\caption{Doppler shifts in the prominence  along the IRIS slit 1 in the wide $\lambda$-band in the  Mg II k line at 12:22:23 UT (solid line) and in a narrow $\lambda$-band in  H$\alpha$ (dashed lines) for two cuts close to  the IRIS slit position according to the accuracy of the co-alignment between the MSDP and IRIS. Positive/negative  values are red/blueshfits.}
 \label{cuts}
 \end{figure}

%
\begin{figure*} 
\centerline{\includegraphics[width=0.8\textwidth,clip=]{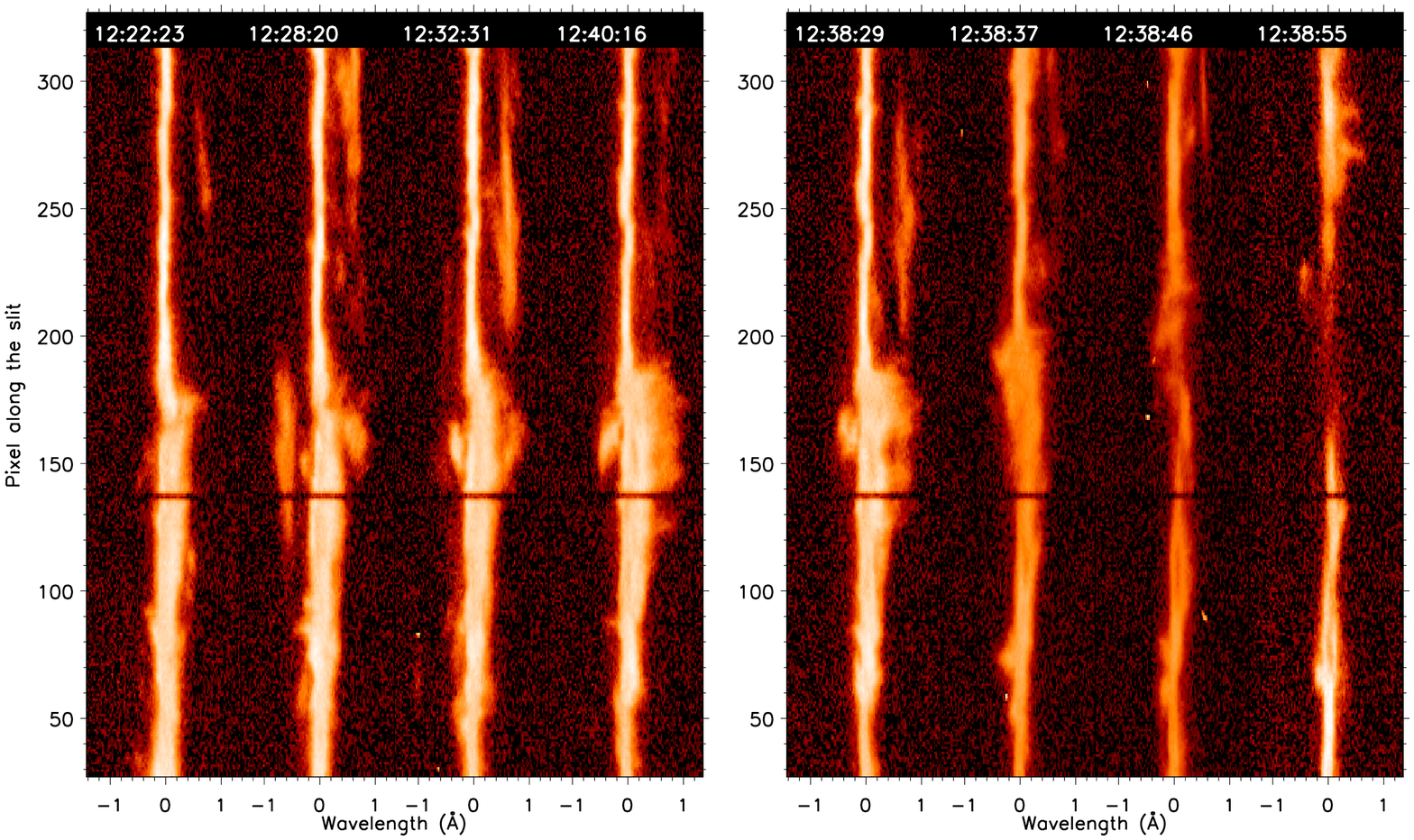}}
 \caption{Examples of Mg II k line spectra:   evolution of the Mg II 2796.4   \AA\ spectra  at 12:22, 12:28, 12:32, 12:40 UT  in  slit 1 ({\it left}).  Examples of 2796.4 \AA\ spectra in the four slits of IRIS located from east to west in the slit jaws (slit 1 at 12:38:29, slit 2 at 12:38:37 UT, slit 3 at 12:38:46 UT, slit 4 at 12:38:55 UT)  ({\it right)},
The spectra  are between pixel 8 and   pixel 314 along the slit. The wavelength range of the Mg II spectra is 2795.00 - 2797.78 \AA\ and the length of the slit is 47 ''.  The dark line at pixel=136 is an instrumental feature used  for the alignment of the different  lines in the spectra. The zero is taken as the velocity reference. North is up. {\bf An online movie  of the Mg II k spectra  is available, see  Figure \ref{movie} in the Appendix }.
}
 \label{iris1}
 \end{figure*}

\subsection{Dynamics of the Ca II prominence}

The Ca II line movie (Ca II movie online) reveals  turbulent behavior of the plasma  in different parts of the prominence
 (top box in Figure \ref{msdp2} {\it top right panel)}. It is impossible to follow  the evolution of structures. The plasma is fuzzy. This part corresponds to the northern  part of  the IRIS slits.
In the middle of the IRIS slits, the Ca II images show very clearly a   small  dark bubble  (3 arcsec) surrounded by bright rims (middle box in Figure \ref{msdp2} {\it top right panel)}. It is also visible in the Mg II lines.  This bubble corresponds to 
 pixels in the spectra where there is a change in  behavior of profiles: from narrow to broad  with double peaks. In the bright rims knots go  down frequently around 13:00 UT.  In the left part on the images we  also see oblique threads striating the prominence. These threads exhibit counter streaming. 
 Finally we see the emergence of a bubble (semi-sphere of 20  arcsec of diameter) close to the limb in the right part of the prominence (bottom box in Figure \ref{msdp2} {\it top right panel)}. Between 12:10 UT and 13:30 UT, the bubble rises  to an altitude of 10 arcsec  over the limb with a velocity of 2 km/s. It looks similar to the bubbles previously observed with Hinode/SOT  by \citet{Berger2011,Dudik2012}.  The authors invoked  an emerging flux as the cause of such bubbles in a weak  bipolar field environment.  Thermal instability  \citep{Berger2011}  or magnetic pressure excess \citep{Dudik2012} would lead to the  slow  rise  of   bubbles through  the prominence.
The location of the large bubble  in the Hinode FOV  is  unfortunately  outside of the IRIS FOV. No  study on the temperature and the density of the bubble can be achieved  using  the IRIS spectra to resolve the question of the existence of a thermal instability.  We suggest that the magnetic solution is acceptable and the dark area in the bubble is the corona that we see under the prominence. Slow rising prominence  before eruption is a common  observation.
  Magneto-hydrodynamic  models based on the torus instability tell us that a prominence modeled as a flux rope should reach a given threshold height to  erupt     \citep{Gosain2012,Torok2009}.

\subsection{H$\alpha$ profiles}
We obtained a map of H$\alpha$ profiles on the whole  prominence for each MSDP observation time. From the line  profiles,  Doppler shifts have been computed by the bisector method for H$\alpha$ +/- 0.3 \AA. 
We focus our study on the time 12:22 UT (Figure \ref{msdp2}).
The range of the values is  between - 3  km/s and 4.5 km/s, consistent with earlier observations of prominences \citep{Labrosse2010}. Large Doppler shifts are only measured at the tops of prominences where only  a few threads are integrated along the line of sight \citep{Schmieder2010}. 
High velocity threads in H$\alpha$  larger than 20 km/s cannot be detected in these observations because  the  H$\alpha$ wavelength band  of  the MSDP is too narrow.
The evolution of the Doppler shift pattern  is relatively fast  (Figure \ref{msdp3}).

We present in Figure \ref{cuts} the  cuts of the H$\alpha$  Doppler shifts along  the  slit  1  of IRIS FOV at 12:22:23UT (two  possible positions according to the accuracy of the co-alignment). The intensity of the H$\alpha$ line along the slits 2, 3,  and 4 is  under the threshold of the intensity where   Dopplershifts can be computed.  The Doppler shift structures have a size on the order of 18 arcsec. The H$\alpha$ Doppler shift values  are lower than those of the Mg II lines, which can be explained by the lower spatial resolution of the MSDP and by the fact that the H$\alpha$ line has lower optical depth so that the observations integrate more structures along the line of sight (LOS).  The seeing  also smooths the values.


\begin{table*}
\caption{Characteristic parameters of the Mg II line profiles observed by IRIS (observation) and copied in \citet{Paletou1993} (model).}
\begin{tabular}{llll|ll}
\hline
parameter&unit& Observation &(IRIS)& Model  of & \citet{Paletou1993} \\
&&Mg II h& Mg II k&Mg II h& Mg II k\\
\hline
FHWM&\AA&0.15&0.16&0.40&0.40\\
Central intensity&10$^{-7}$ erg s$^{-1}$ cm$^{-2}$ sr$^{-1}$  Hz$^{-1}$&2-2.4&2.3-3.1&1- 4&1- 4.3 \\
Integrated intensity&10$^{4}$ erg s$^{-1}$  cm$^{-2}$ sr$^{-1}$&1.2-2.6&1.8-3.7&2.4&3.7\\
\hline
\end{tabular}
\end{table*}

\begin{table*}
\caption{Several components of Mg II  line profiles at 12:28  UT  and at 12:38 UT  for some pixels along the slit (Dopplershift velocity  in km/s).}
\begin{tabular}{llllll}
\hline 
pixel number &peak1&peak2&peak3&peak4&peak5\\
\hline
140 &  -64.2&& -4.1&24.5&\\
150 &-67.0 &-29.0&0.6&25.5&75.6\\
160& -63.8&&-1.1&20.6&67.5\\
200&&& -1.5&&\\
250& &&0.4&&62.9\\
300&&&2.3&42.4&65.1\\
\hline
100&&&-5.0&15.5&\\
164&-41.9&-22.7&2.8&32.8&72.1\\
270&&&0.7&&71.5\\
\end{tabular}
\end{table*}

\begin{figure*} 
 \centerline{\includegraphics[width=0.9\textwidth,clip=]{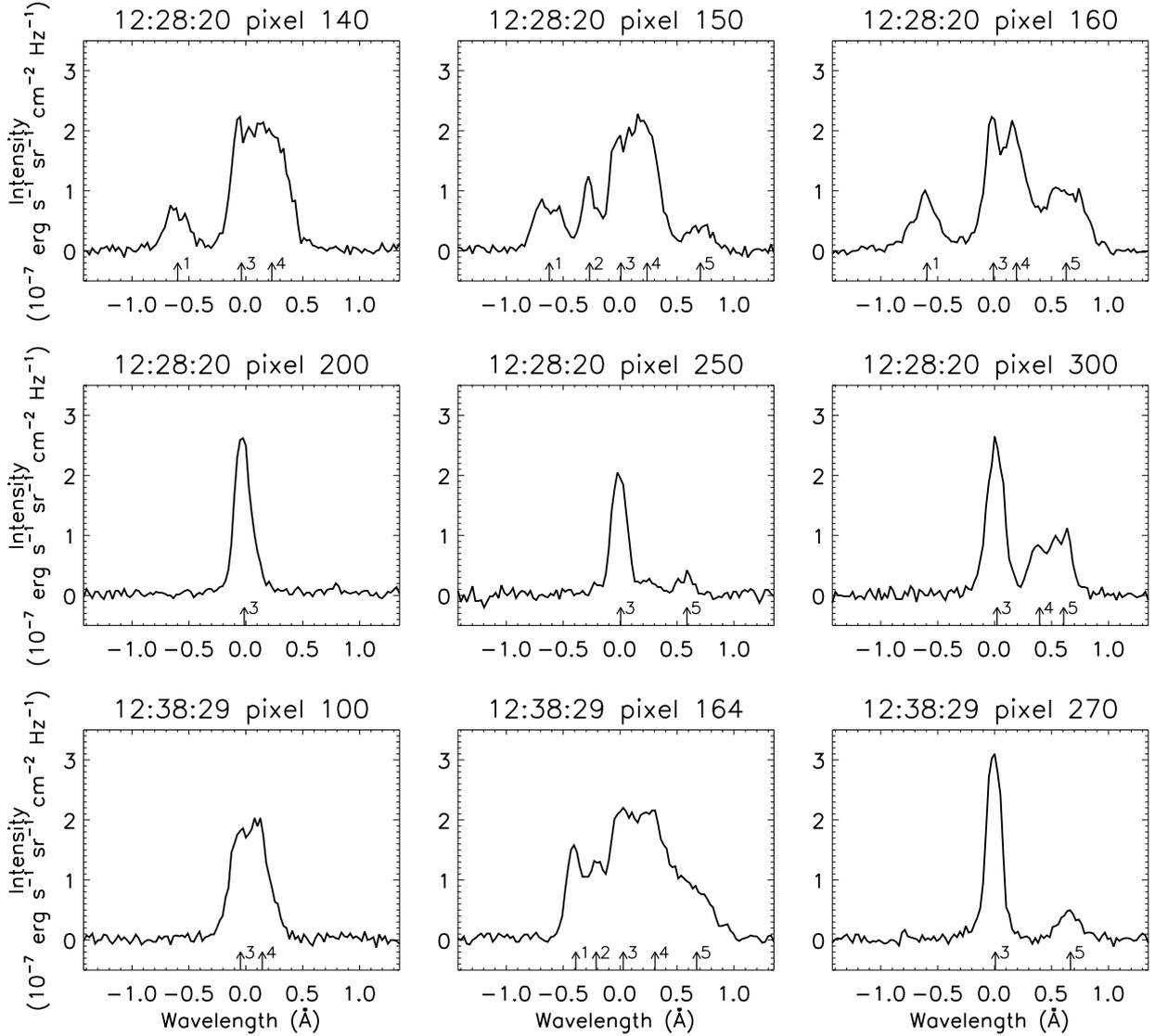}}
 \caption{IRIS line profiles of Mg II 2796.4~\AA: ({\it first and second rows}) at 12:28:20 UT for six pixels along the slit (140, 150, 160, 200, 250, 300);  
({\it third row})  at 12:38:29 UT for y=100, 164, and 270 (see spectra in Figure \ref{iris1}). 
We note  that lateral peaks  (indicated by ticks)  are resolved   on both sides of the main peak indicating fine structures along the LOS with Doppler shifts reaching 60 to 80 km/s (see Table II). }
 \label{iris2}
 \end{figure*}

%
\begin{figure*} 
\centerline{\includegraphics[width=0.5\textwidth,clip=]{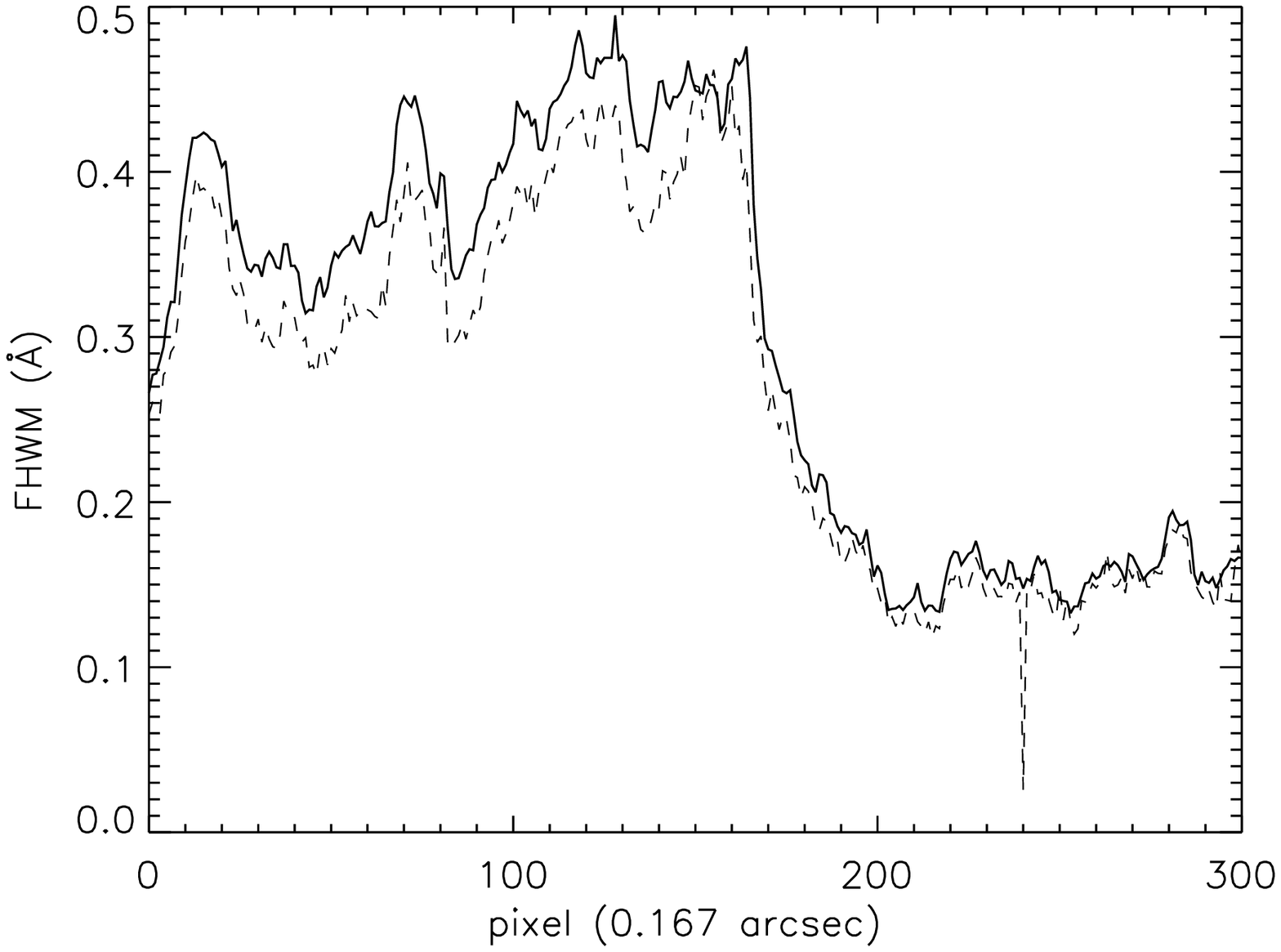}
\includegraphics[width=0.5\textwidth,clip=]{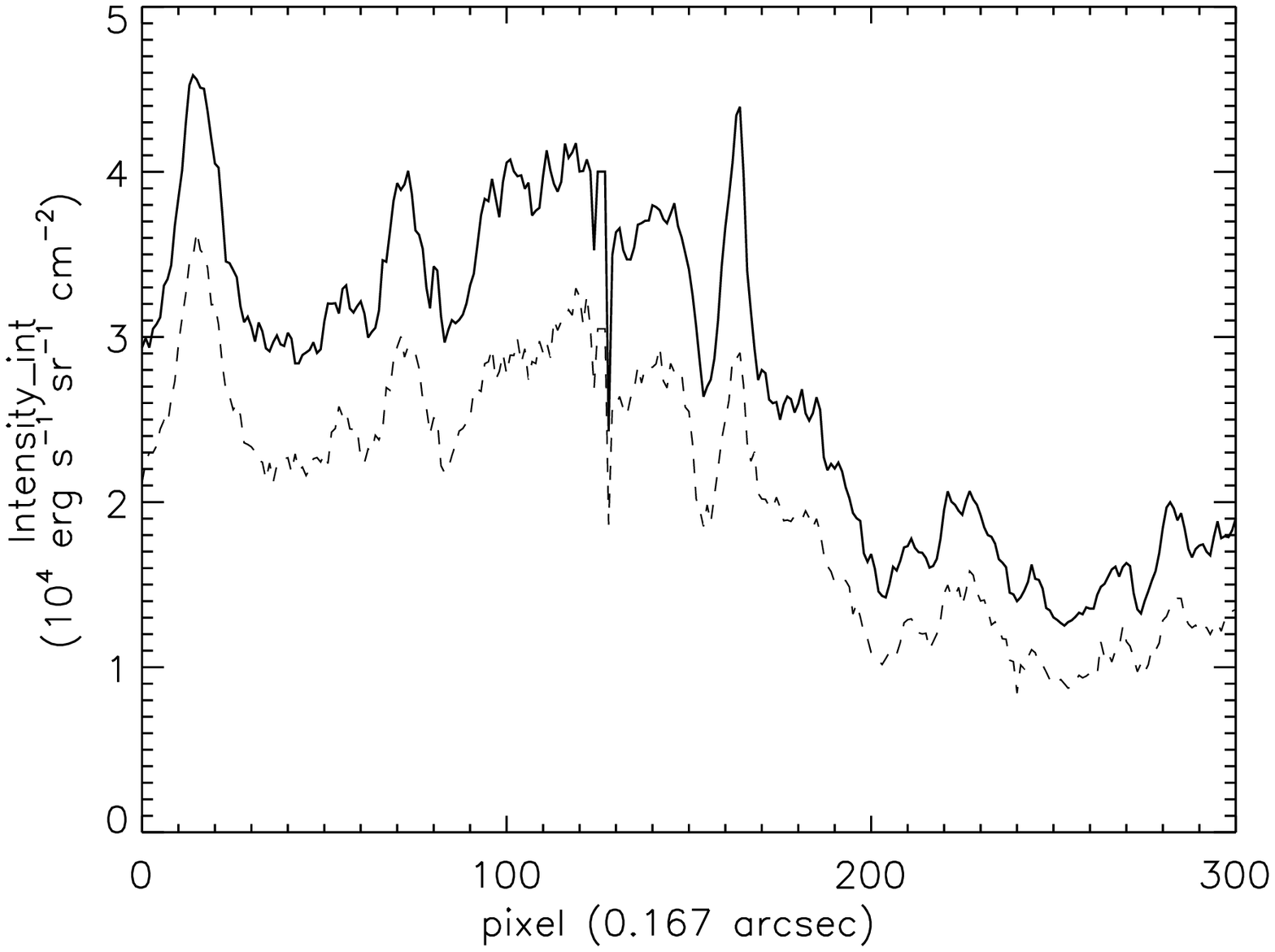}}
\centerline{\includegraphics[width=0.5\textwidth,clip=]{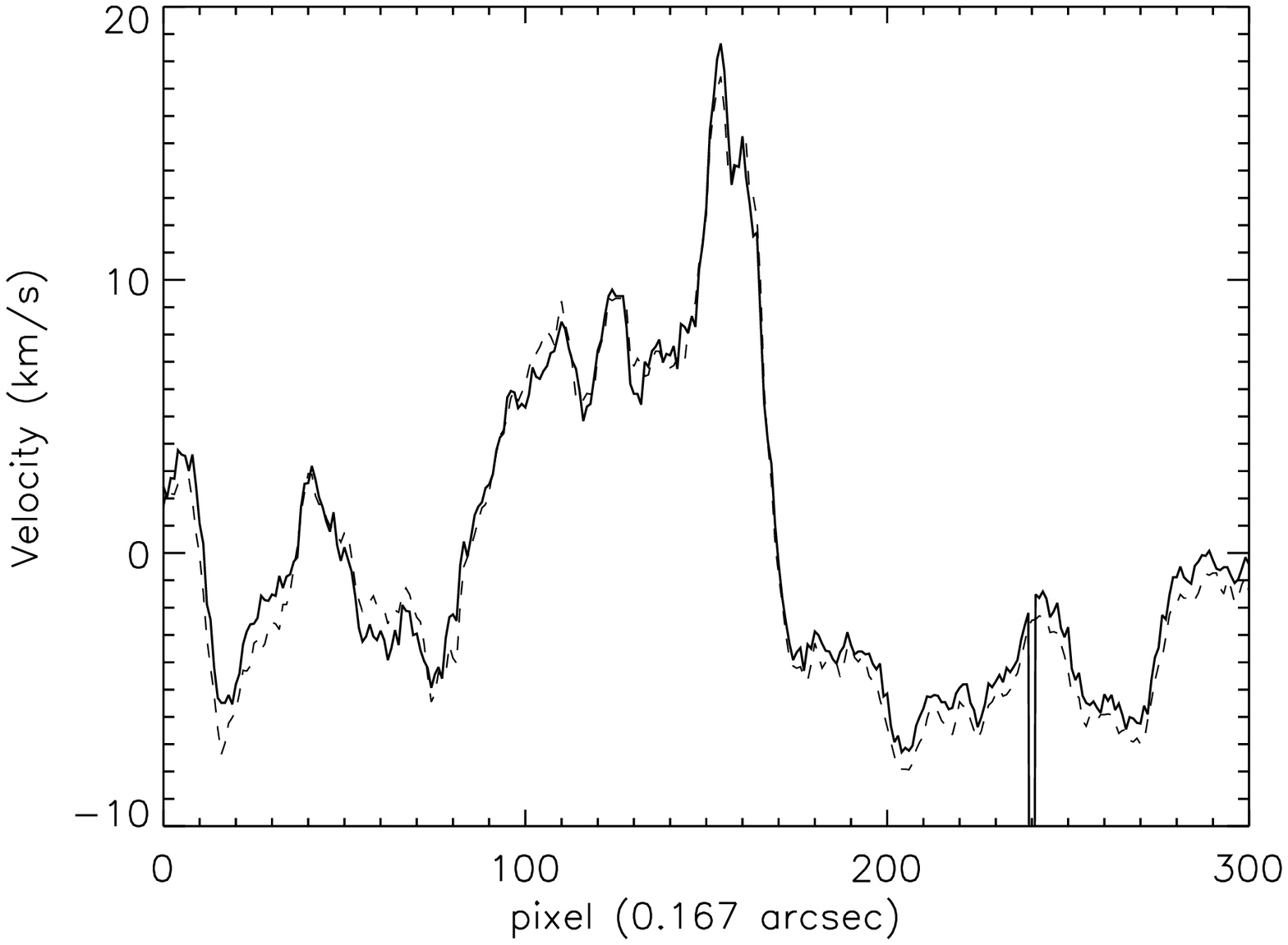}
\includegraphics[width=0.5\textwidth,clip=]{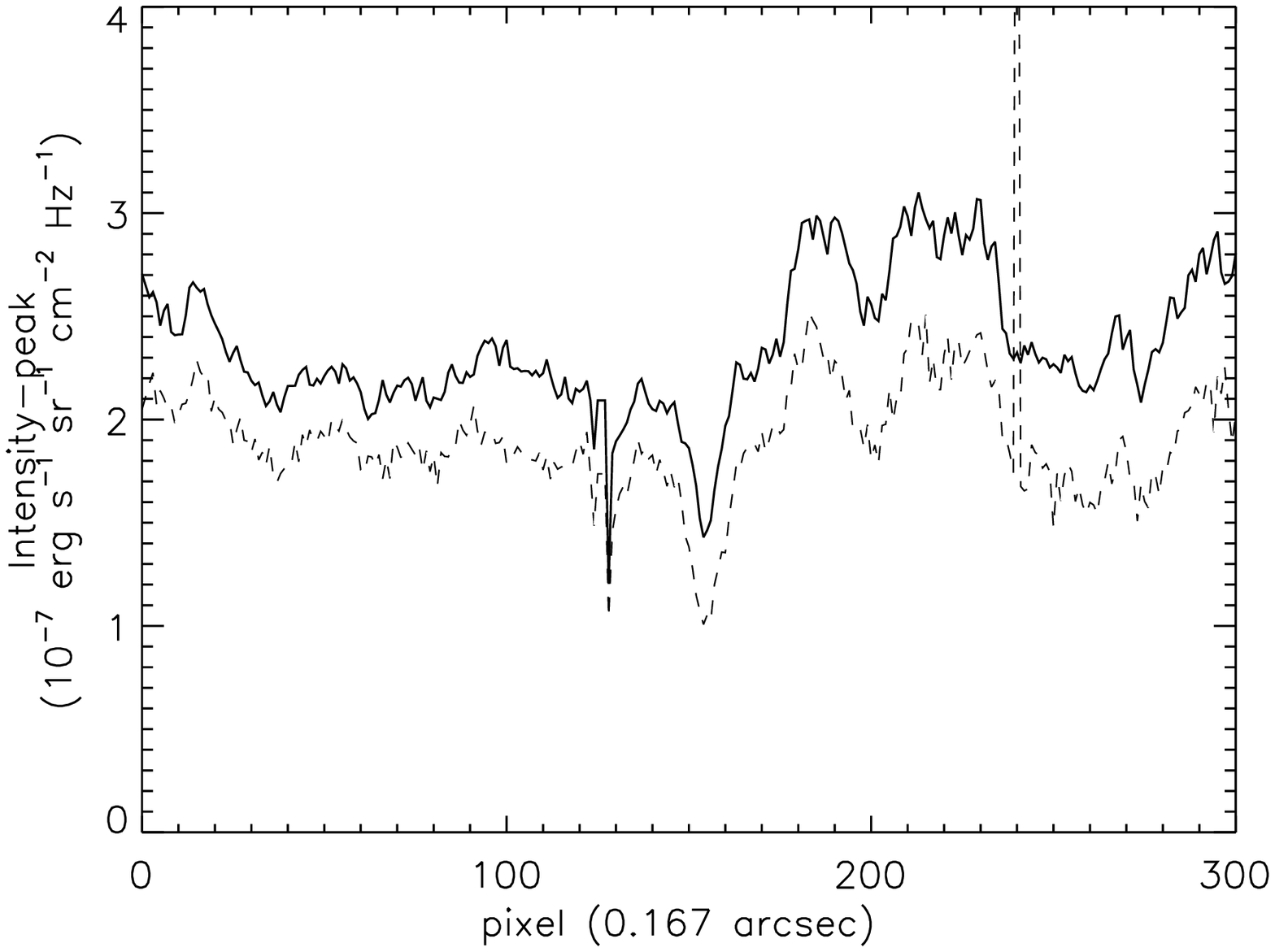}}
 \caption{Characteristics of the Mg II  k 2796~\AA\ (solid line) and Mg II  h 2803~\AA\  (dashed line)  along the IRIS slit 1 at 12:22:23 UT: full half width maximum (FHWM), integrated intensity (I-int), Dopplershifts (velocity), and  peak intensity (I-peak)  assuming Gaussian profiles.
 The ratio of the peak intensity between the two lines is equal to 1.24.}
 \label{mg2}
 \end{figure*}

\subsection{Mg II profiles}
The IRIS profiles are very interesting. In the northern part of the slits the profiles are very narrow; in the southern part the profiles have secondary peaks (Figs \ref{iris1} and \ref{iris2}, IRIS movie online).  The spectra correspond to pixels between 8 and  314 along the slit in the slit jaws with a pixel size of 0.167 arcsec.  Horizontal dark lanes are the horizontal fiducial lines
used for co-alignment between the different wavelengths.  Slit 4  corresponds to the extreme  west edge of the triangle-shaped prominence in the slit jaw and is frequently empty of signal or with a  long gap of very weak signal.    We focus our study mainly on slit 1 located at the east in the slit jaw where the  signal is the strongest.
The  dispersion per pixel $\Delta\lambda$   is equal to  0.02546 \AA\  and  $\lambda_0$=55.47 pixel is the centroid of the averaged line used as the rest wavelength.
The most of the profiles of the Mg II lines  are not reversed and the maximum peak intensity is  around  3 x 10$^{-7}$ erg s$^{-1}$  sr$^{-1}$ cm$^{-2}$ Hz$^{-1}$ (Table I). The FHWM of the Mg II lines is around 0.15 \AA . If we fit the whole profile with a single Gaussian   including   the different peaks  we obtained a FHWM of  0.4 \AA.  This is what can be seen with a spectrograph with low spectral resolution \citep{Vial1982}.

We compare 
the observed profiles with the theoretical Mg II line profiles computed  with  non-LTE radiative transfer   models in a 2D static  slab   \citep{Paletou1993} (Table I).  For low peak intensities (lower than 2 x 10$^{-7}$ erg s$^{-1}$  sr$^{-1}$ cm$^{-2}$ Hz$^{-1}$) both  the computed profiles  and  the observed ones are not reversed. For higher intensities the  computed profiles are always reversed. This  is not the case with our  observations. When the observed profiles are wider, we find that there are two structures:  one  is nearly static and the other one has a velocity on the order of 20 km/s (Table II). Only a few line profiles are reversed.  We consider reversed profiles when the two peak wavelength positions  are more or less symmetric with respect to the line center (for example, in Table II at y=100 at 12:38 UT and perhaps y=160 at 12:28 UT, but the peak positions  are  not symmetric there). The distance between the two peaks in the  case of reversed profiles is around 0.2 \AA.
The ratio between the maximum intensity of the two Mg II lines (k/h) is around 1.24;  between the integrated intensity it is 1.33,   the FHMW maximum ratio is similar (1.1),  and the Doppler shifts are equivalent (Figure \ref{mg2}).

The predicted computed  profiles do not fit the observations in many respects. Similar surprising results were  found when we observed the hydrogen Lyman line series in prominences. Some prominences     presented  profiles that were not systematically reversed \citep{Heinzel2005,Schmieder2007,Gunar2007,Curdt2010,Schwartz2012}. This has been discussed in terms of the orientation of the magnetic structures in prominences and the column mass.

The narrow profiles in the northern part of the slit correspond to small Doppler shifts ($\pm$ 5 km/s). This part is in the section of the prominence plasma that looks  turbulent  in the Ca II movie.  For unresolved turbulent plasma, the profiles should be broaden. This  is not the case. 
The secondary weak peaks of the profiles  are wider and have a long extension along the slit (e.g., at 12:32:32~UT in Figure \ref{iris1}). They could correspond to  small scale turbulence, but these  peaks are mainly redshifted and not symmetric versus the line center as we expect for unresolved  turbulence. The global structure with turbulence plasma  would be    globally redshifted in that case.  

%
\begin{figure*} 
\centerline{\includegraphics[width=0.7\textwidth,clip=]{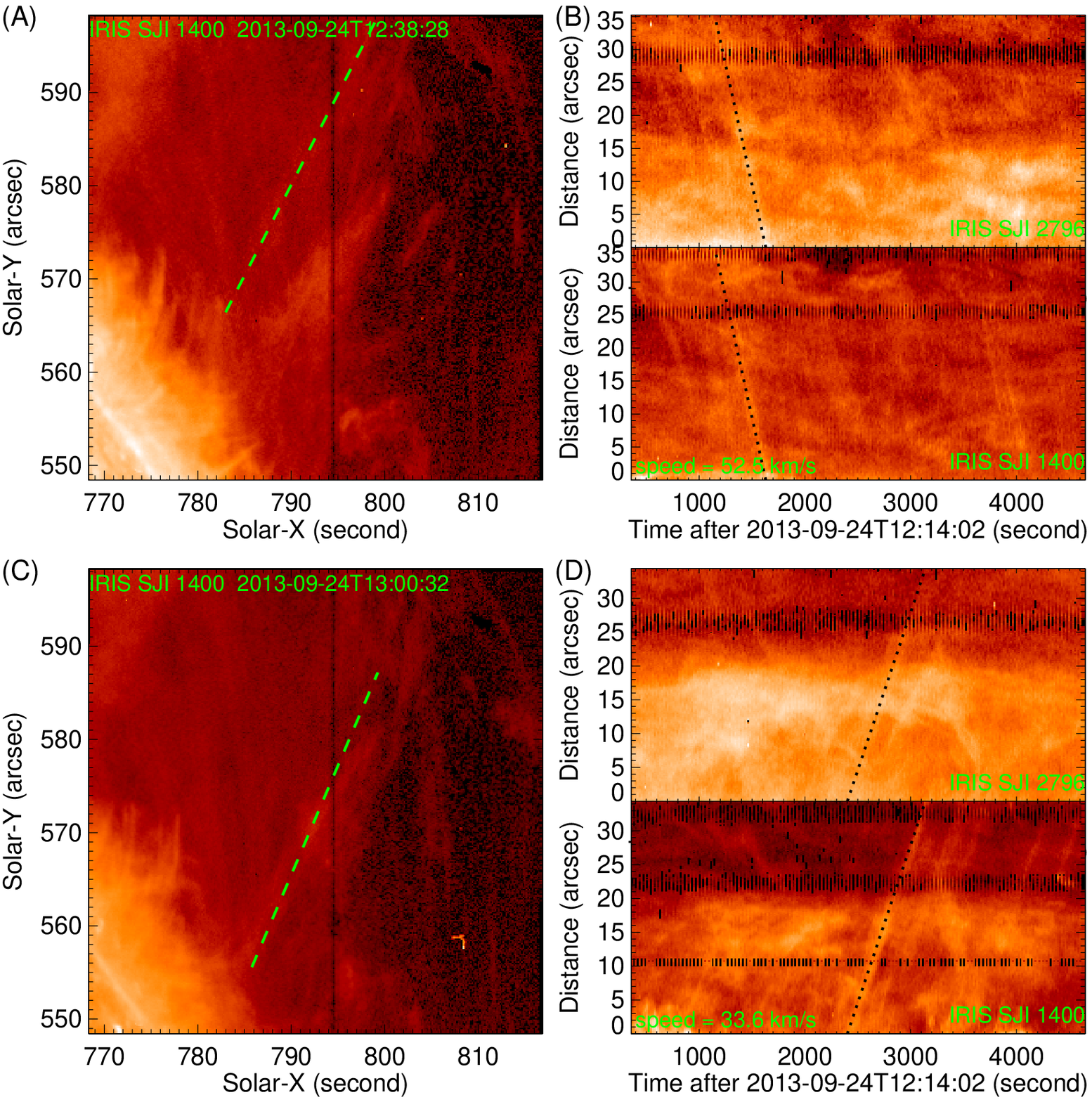}}
\caption{({\it left panels}) (A) and (C )  Si IV slit jaws  (47'' $\times$ 47'' ) respectively  at  12:38:28 UT  and 13:00:32UT showing the cuts  of the time slice when the flows are crossing the IRIS slits. ({\it right  panels}) The time slice diagrams are presented in  pairs corresponding to MgII and Si IV  slit jaws  for each cut. (B) corresponds to the cut in (A), (D) to the cut  in ( C). The time slice diagrams  show parallel  straight lines indicating flows going towards  the southeast/northwest direction with a projected  speed respectively  of 52.5 and 33.6 km/s.
 (a Si IV slit-jaw  movie is available in the  online version  with  Figure \ref{movie} in  the Appendix  explaining the movie.  )
 }
 \label{time_iris}
 \end{figure*}

%
\begin{figure*} 
\centerline{\includegraphics[width=0.7\textwidth,clip=]{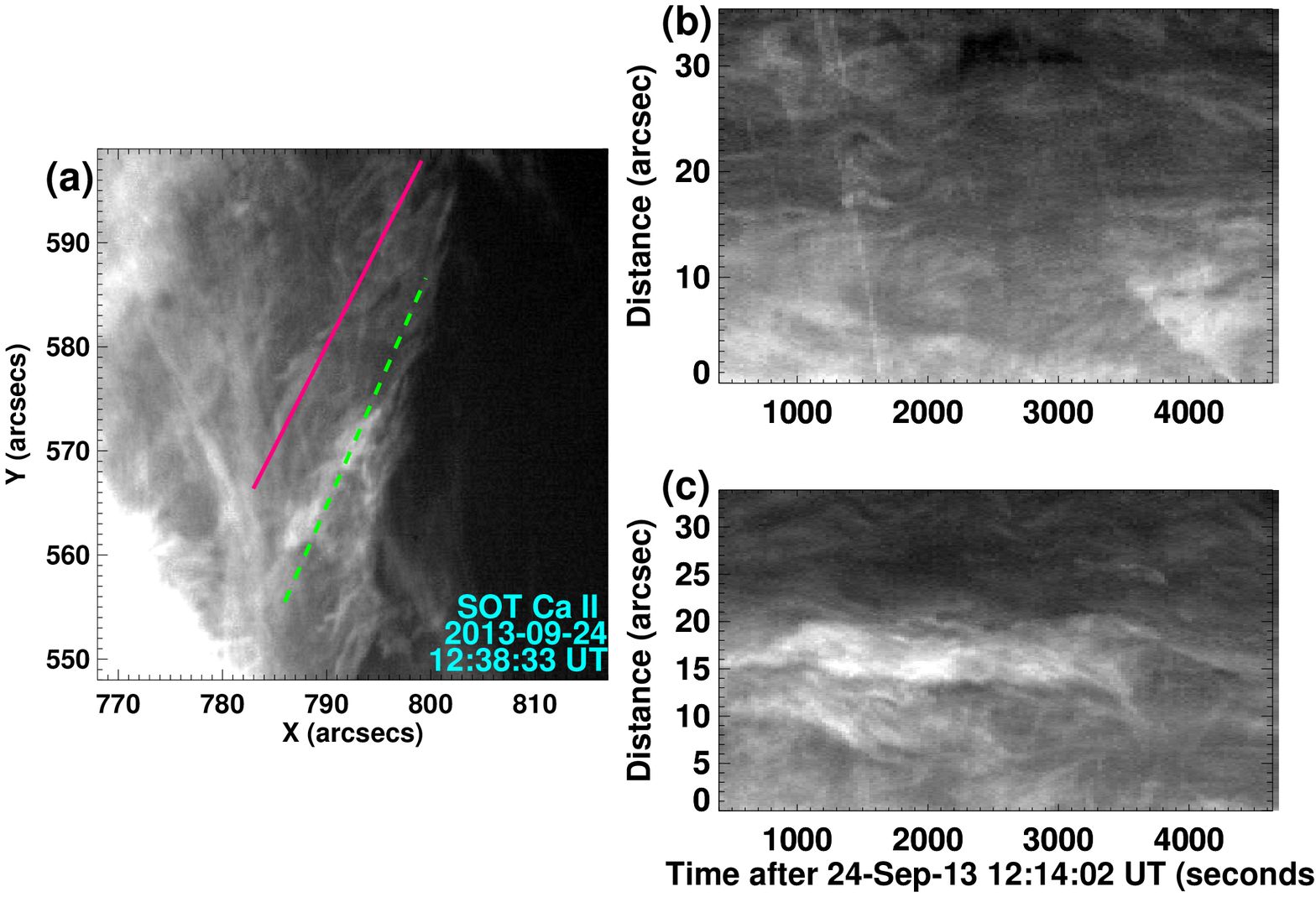}}
\caption{(a) Ca II image showing the two cuts used to make the time slice diagrams presented in ( c) and (d). 
The time slice diagram in ( c)   shows parallel  straight lines indicating flows going towards  the southeast/northwest direction; in (d) we see mainly a quasi static component and very weak straight lines. }
 \label{time_iris1}
 \end{figure*}

\subsection{Mg II Doppler shifts}

Doppler shifts are computed using  single   or multi-gaussian fits to the peaks of Mg II  line profiles. The Mg II Doppler shifts  computed with single fit are comparable to the H$\alpha$ Doppler shifts.  We show an example of such a comparison for 12:22~UT (Figure \ref{cuts}). 
The Doppler shift/intensity   pattern in  the Mg II h and k lines has a periodicity of  about 5 arcsec in width with smaller structures (1.6 arc sec) along the slit. 
The H$\alpha$ Doppler structures have a triple size. 
The  structures are  clearer in Mg II lines because  the higher optical depth of the line than in H$\alpha$ (Heinzel, private communication).  

In the northern part of the slits, the profiles are narrow and  usually one Gaussian profile fits the observations well.
In the southern part,  fine structures in the Mg II lines are observed in the profiles (Figure \ref{iris1}).   The spectra  along the slit  (at 12:28 UT  and at 12:38 UT, for example) show many different structures along the LOS (Figure \ref{iris2}). We used multicomponent Gaussian fits  to compute the Doppler shifts of each individual thread crossing by the LOS (Table II).   A relatively static component is always present, exhibiting the highest peak (peak 3). High Doppler shifts of up to 70 - 80  km/s are detected. In some pixels (y=160), we note the structures with opposite Doppler shifts  suggesting that counter streaming flows are present in the prominence \citep{Zirker1998}. 

We have to mention that here we use the centroid of the Mg II line profiles averaged over the whole region as the rest wavelength, since  neutral lines cannot be used in prominence observations to  absolutely calibrate the wavelength.  Nevertheless  there are  no neutral lines in the present  observed wavelength range. In addition  it has been found that the orbital variation of neutral line positions in other IRIS observations is usually less than 5 km/s peak to peak. So here we estimate an uncertainty of about 5 km/s for the velocity determinations (References in  IRIS technical note 20-ITN 20 Wavelength Calibration-
\url{http://iris.lmsal.com/documents.html}).
This means that regarding the 5 km/s velocity as stationary is reasonable.

\subsection{Transverse flows measured in Mg II and Si IV slit jaws}

With a time slice procedure, we  computed the
transverse velocity field in some moving structures using the slit-jaw images (Figure \ref{time_iris}).    
In the movies (online) of the slit jaws, the global structure changes slowly. However, fine threads exhibiting fast moving material  cross the slit. For example, in pixels around  y= 300  material goes towards the limb between 12:29 UT and   12:39 UT,  and  for pixels  in the middle of the raster
  material   goes to the north  between 12:56 UT and  13:02 UT (Figure \ref{time_iris}). 
The  projected speeds  of these features on the plane of the sky, computed from the time slices   are respectively  52 km/s and   33 km/s.\\

The measured speeds are the same for  the two slit-jaw sets (SJI 1400 and  SJI 2796).
These transverse velocities correspond to knots following threads that  cross  the IRIS slit around 12:38 UT and 13:01 UT.
The transverse velocities  correspond to high Doppler shifts detected in the structures crossing the slit. This  indicates that the fine threads
moving at the front of the prominence are oriented at a given  angle with respect to the plane of the sky. For the knot crossing the slit at 12:38 UT the angle of the thread is around 50 degrees with the plane of the sky.  The velocity vector strength may reach up to 
  100 km/s. These fine threads could be  fine threads  along the spine of the prominence  between A and B, or between A and C (N-S filament in Figures \ref{meudon} and  \ref{msdp2}).

We performed a similar analysis with Ca II images and compare it with IRIS time slice (Figure \ref{time_iris1}).  The brightest pattern  with no significant motions is similar in Ca II and in Mg II. It could be interpreted as   corresponding  to the quasi-stationary plasma. There is also an analogy between the  fine threads crossing the slit, 
but it is not clear if these are  exactly the same threads or  parallel threads.  

%
\begin{figure*} 
\centerline{\includegraphics[width=0.8\textwidth,clip=]{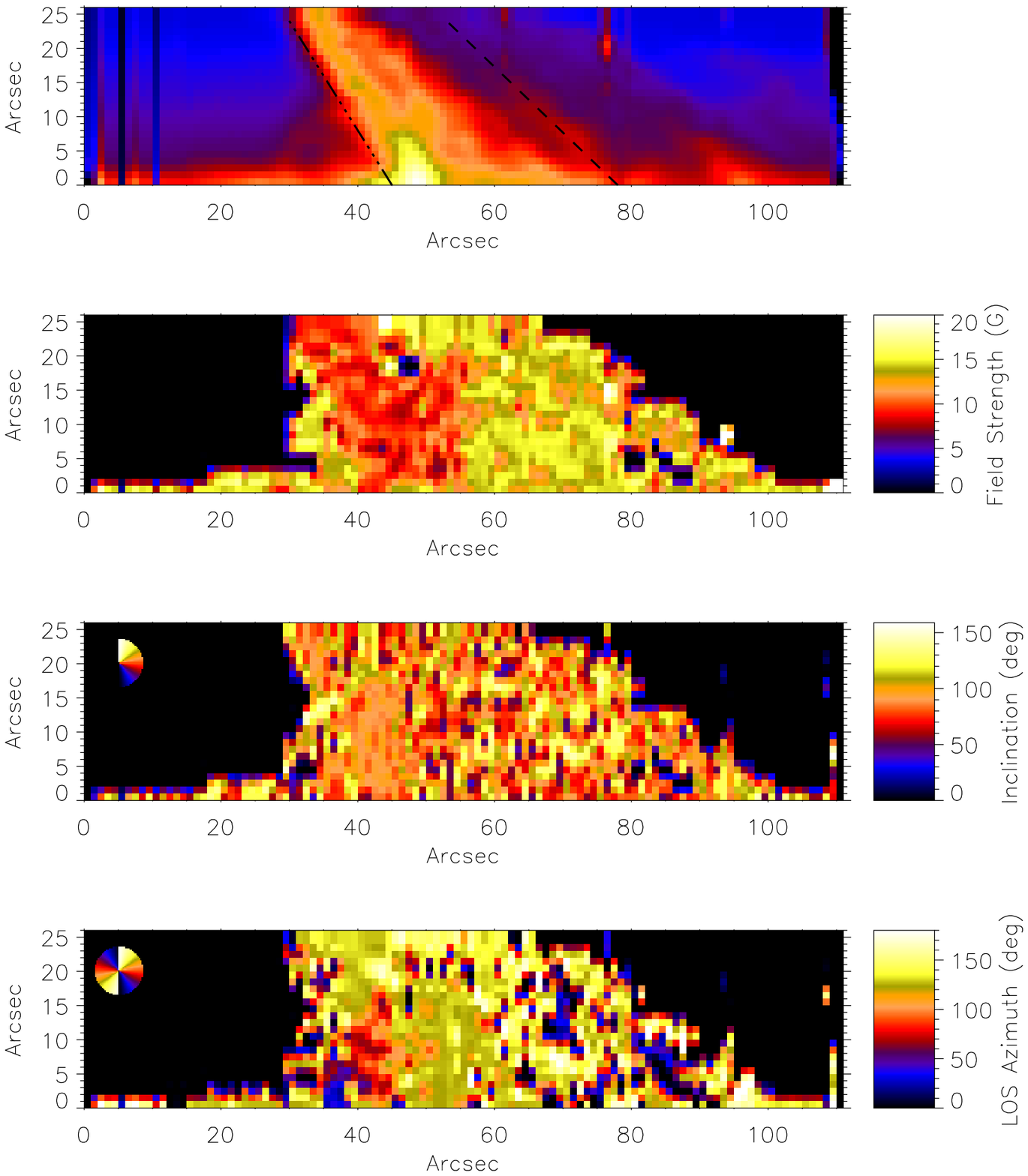}}
 \caption{THEMIS observation on September 24, 2013, obtained between 12:24 UT and 14:30 UT: ({\it from top to bottom})  intensity, magnetic field strength, inclination, and azimuth. The two  lines (dot-dashed and dashed) in the intensity image represent the location of the two  cuts used  in Figure \ref{themis3}. The origin of the cut in the brightest region is at x=40, y=0;  the origin of the IRIS slit is   at x-80, y=0.
 }
 \label{themis2}
  \end{figure*}

\subsection{Magnetic field vector: macroscopic  component}

 The raw data of the THEMIS/MTR mode was reduced with the DeepStokes  procedure \citep{LAARMSDG09} and the Stokes profiles were fed to an inversion code based on 
 Principal Component Analysis \citep{LAC02,Casini03}.  Initially, the observed profiles  were compared against
those in a database generated with known models of the polarization profiles of the He D3 taking into account the Hanle  and  Zeeman effects \citep{LAC02}.  
The details of the MTR data reduction are completely explained in \citet{Schmieder2013}. The most similar profile of the data base containing 90000 profiles is
kept as the solution and the parameters of the model used in its computation 
are kept as the inferred vector magnetic field height above the photosphere and scattering angle. Error bars are determined for those parameters as well by 
doing
statistics on all other models which are sufficiently similar to the observed ones, although not as similar as the one selected as the solution.
  

Figure \ref{themis2}  presents the maps obtained after inversion  of the Stokes parameters recorded in the He D$_3$  line: (a) intensity, (b) magnetic field 
strength (c) inclination, (d) azimuth.  The angle origin  of inclination is the local vertical, the  origin of the azimuth is  the LOS  in  a 
plane containing the LOS  and the local
vertical. We see that the brightest parts of the prominence have a mean  inclination of 90$^\circ$ which means that the magnetic field  in the brightest  
oblique structure  is clearly horizontal.  However, there is a large dispersion of the values (+/-30 degrees) from one pixel to the next in the lateral parts of 
the prominence.

\begin{figure*} 
\centerline{\includegraphics[width=0.6\textwidth,clip=]{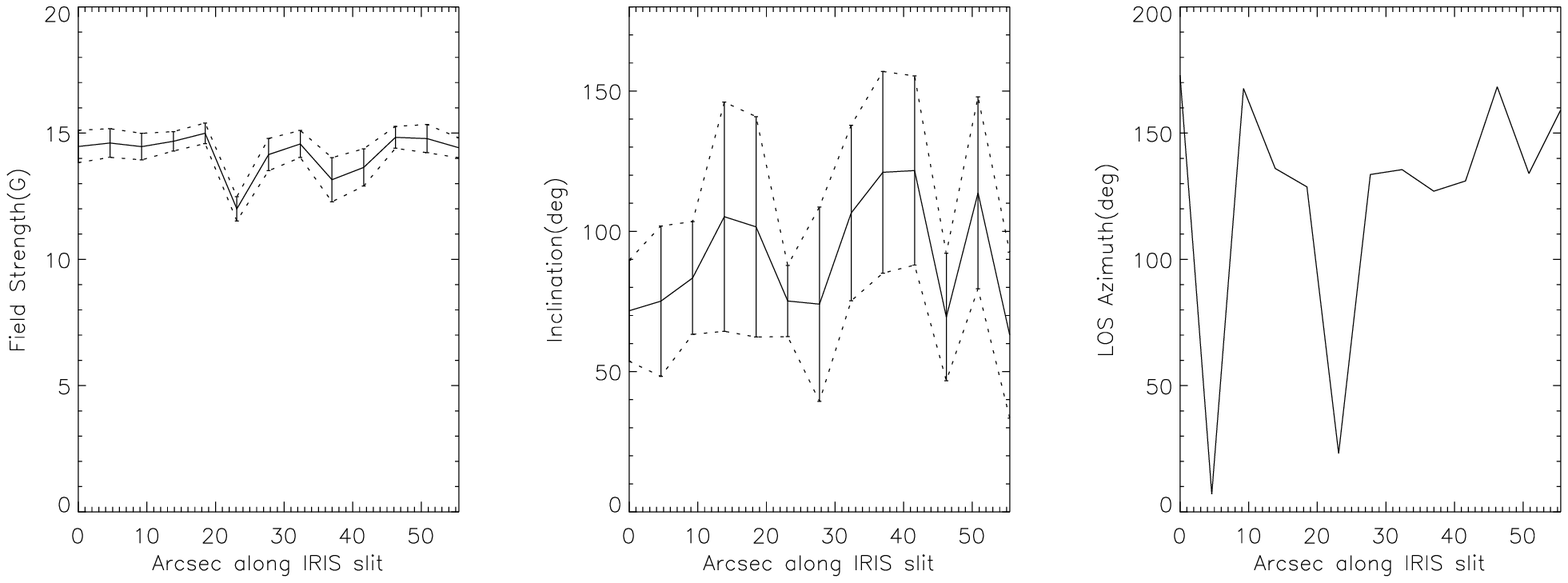}}
\centerline{\includegraphics[width=0.6\textwidth,clip=]{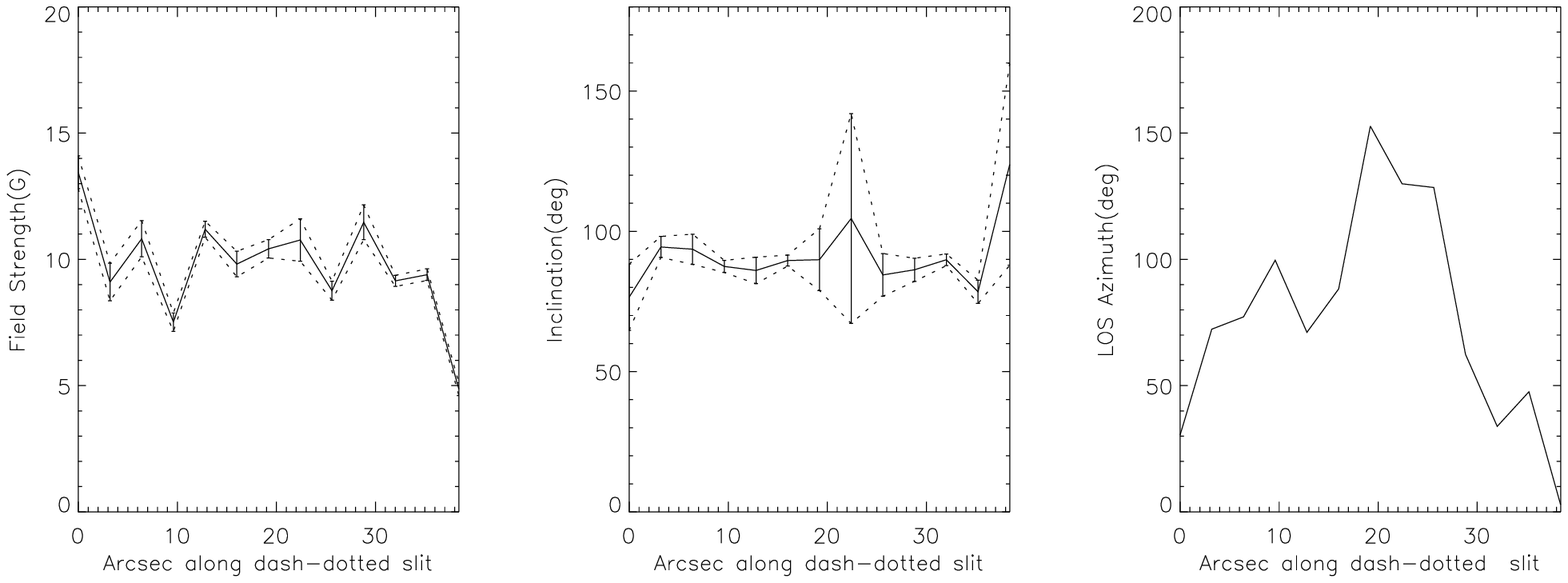}}
\caption{Variation of the magnetic field, strength,  the inclination, and the azimuth (THEMIS): ({\it top panels}) data along the IRIS slit,  
({\it bottom  panels}) data along  the bright  structure.}
\label{themis3}
\end{figure*}

 Figure \ref{themis3}   presents the  variation of magnetic field strength, inclination, and azimuth along the  slit positions of IRIS in Figure \ref{themis1}. 
 The field strength   is  in the range 5-15  Gauss and mainly  horizontal.  The inclination is around 90 $^\circ$ +/-30$^\circ$. The azimuth is  close to  
 90$^\circ$ and again with a large dispersion (+/- 50 degrees). This means that the magnetic field vector is mainly perpendicular to the plane of the sky with  
 a  large dispersion of values.

 The brightest part of the prominence where the magnetic field  is directed horizontally with respect to the solar surface is located mainly in foot A at the 
 intersection of the two sections of the filament (Figure \ref{meudon}).
  This confirms previous results \citep{bommier1998,Casini03}. Prominence feet 
 observed on the disk have also shown that the field lines are tangent to the photosphere  \citep{lopez2006,Schmieder2013}.  Their shapes  have been  
reliably represented  by  
 linear force free field extrapolations \citep{aulanier1998,Dudik2008,Dudik2012}.  
 The dispersion of the values of the inclination and the azimuth could be due  to the  foreground  transient structures of  the  large arch seen  in 
 304  \AA\, (Figure \ref{themis1}).

%

\subsection{Magnetic field: turbulent component}

The model used for inversion, and whose results we have just discussed, assumes a unique magnetic field vector per pixel. The low signal-to-noise ratio of the
 polarization signature advise against assuming more complex scenarios. This single value of the magnetic field vector can be seen as the large scale magnetic
 structure supporting the prominence and revealed by polarimetry at low spatial resolutions.  In the previous subsections we have listed a compilation
 of observations from IRIS and Hinode SOT with high cadence and high spatial resolution that suggest a more complex magnetic scenario where the local magnetic 
 structure departs from the stand-alone macroscopic horizontal field. 
 
Intrigued by the possibility of unveiling a more dynamic and complex field structure in the prominence
we explored the effects of a turbulent field added to a macroscopic 
field into the Stokes profiles of the He D$_3$ line. We should say that the word {\em turbulence} is used incorrectly here. It may mean actual turbulence
in the sense of hydrodynamics, but it may well just mean several unresolved (in time, space, and along the line of sight) magnetic components that are added 
together in the same pixel of our data set. 
The polarization of the He D$_3$ line in the 
presence of an isotropic field with a strength of roughly 15~G can be readily computed and, to 
a good precision, results in the absence of Stokes U and V profiles, while every transition 
involved in the line formation is polarized in Stokes Q to $\frac{1}{5}$ of the polarization at 
zero field \citep{Landi1984}. The absence of Stokes U and V profiles in a turbulent field already excludes this possibility from our observations: we do see
Stokes U and V signals. If there is a turbulent field at all, it is mixed with a macroscopic average field.
Therefore, the  single profile in Stokes Q emitted by the turbulent magnetic component was added to the profiles from the macroscopic magnetic component
weighted by a filling  factor.  These  new profiles made of the addition of the two components were inverted with the same model of the 
observed data. The solutions found have systematically smaller field strengths and the 
inclination is on average the same, but the error bars grow enormously. 
This growth in the 
error bars of the inclination is exactly what is observed in the region sampled by the IRIS slit. 
The spectropolarimetry of the D$_3$ line would therefore be consistent with a turbulent 
component as suggested by the other observables cited above and added to a macroscopic field. The field   
strength in Fig.~\ref{themis3} (top panels) would be just a lower limit to both the macroscopic and the turbulent components, while the inclination
of the macroscopic component would still be horizontal. 
On the other hand, those parts of the prominence with small error bars in the inclination, like 
those in the bottom  panels in 
 Fig.~\ref{themis3} across the brightest parts of the 
prominence, would not accept a turbulent component but just one single macroscopic field. In 
this picture the prominence would be made of an organized horizontal and relatively weak field 
supporting the densest cores of the plasma plus some other regions with stronger horizontal 
fields in addition to a turbulent field responsible for most of the rapid dynamics of the plasma.

\section{ Discussion and conclusion}
A large  quiescent prominence on the northwest limb was the target of coordinated observations  on September 24, 2013.  Observed on the disk a few days before, we note that the filament consisted of two sections: one oriented east-west  and one oriented north-south.
 The multiwavelength observations of this prominence obtained with  the new spectrograph IRIS,  Hinode/SOT, SDO/AIA,  and from the ground (THEMIS in Canary Islands and the MSDP spectrograph on the Meudon solar tower) were analyzed.  The small field instruments (109 arcsec $\times$109 arcsec for the Hinode/SOT, 50 arcsec $\times$ 50 sec sec for IRIS) were focused on the junction of the two sections: the  foot A and the integration of the EW  filament section.  The prominence is shaped like  a  triangle.  The AIA and the MSDP with their large fields of view  also show  the section NS with different arches and feet. We note the different aspects of the triangle-shaped prominence  observed in He lines (304 \AA\, and D$_3$) and in the chromospheric lines (CaII and H$\alpha$, Mg II).  This results  from the different temperatures of  formation of the lines but also of their different  optical thickness.

The principal results concern the dynamics and the magnetic field measured in this triangle.   The spectro-polarimetry  of the prominence indicates that the  
magnetic field is mainly horizontal all over the observed region with a  field strength of  only  5 to 15 Gauss.  
This confirms the previous magnetic modeling of filaments  using linear or nonlinear force free field extrapolation showing that prominence material is 
sustained in shallow dips of field lines even in the barb or feet of the filament \citep{aulanier1998,Aulanier2002,vanballegooijen2004,Dudik2008}. The aspect 
of apparent  vertical structures could be just a perspective view of the dips as  showed the simulation of \citet{Dudik2012}  inserting  parasitic polarities in a  shear bipolar 
region  to create a filament.  The analysis of a different prominence by THEMIS had  recently clearly shown the horizontal dips in the feet  of the 
prominence \citep{Schmieder2013}.

Such a stable magnetic field is somehow disturbed by the  dynamics of the observed plasma. First, the observations in 
 the Ca II  movie  revealed a complex tangling of dynamical structures in the prominence. This aspect gives {\it a priori} the same impression as the tangled 
 model for prominence proposed by \citet{vanballegooijen2010}.  We detect some 
  relatively stable  background emission while some threads  in the front have material flowing rapidly. This was also visible in the IRIS slit-jaw movies.  The
  transverse velocities of the fast moving features  along the oblique threads crossing the IRIS slits were obtained with a time-slice analysis. The features 
  or blobs  were running with velocity up to 50 km/s.
  In one part of the Ca II images and in  both of the IRIS slit jaws we see some disorganized motions, which could be interpreted as MHD turbulence.
  
  We explored the possibility of a departure of the magnetic field from the average horizontal field retrieved by our inversions. We excluded a fully turbulent 
  field since it would result in null Stokes U signatures contrary to the observations. However a model made of  a horizontal macroscopic field plus a turbulent
  component would be interpreted by our inversion codes as a horizontal field with large error bars and smaller field strength. This is compatible with our 
  observations in the most dynamic parts of the prominences, those observed with IRIS.
  Our conclusion draws a picture of the prominence where an organized horizontal and relatively weak field 
supports the densest cores of the plasma while some other regions with stronger horizontal 
fields in addition to a turbulent field would be responsible for most of the rapid dynamics of the plasma.

  These rapid dynamics of the plasma was seen at its best in the  IRIS spectra of the Mg II lines.  The Mg II spectra exhibit multiple structures along the 
  line of sight.  The profiles of the lines have a Gaussian profile with weak intensities  and are  not reversed profiles   contrary to the predicted theoretical profiles \citep{Paletou1993}. 
We compute the Doppler shifts  and found discrete values from a  quasi-static  component  ($\pm5$ km/s similar to the  H$\alpha$ Doppler shifts) to 60- 70 km/s.  
The high Doppler shifts correspond to the large transverse motions measured along the oblique threads crossing the slits obtained with a time-slice 
analysis.
 In some pixels  high positive and negative flows are detected suggesting structures  with opposite Doppler shifts crossing the line of sight.  These 
 structures have an angle with the plane of the sky of about 45 degrees, and their  real velocity may reach 100 km/s. Spectroscopy is a powerful  
 diagnostics tool to detect the real orientation of the structures and the real velocities.
 
These oblique threads may be found along the  field lines in the large arches observed in 304 \AA.
Is it  fast moving  cool material  due to fast cooling of EUV plasma? There is counter streaming along these features. 
  It has been suggested that counter streaming could occur because of   longitudinal oscillations \citep{chen2014,Luna2014}. The material in the dips could 
  travel on one side or the other side. For large scale structures the siphon flow mechanism could be more important than the counter streaming leading to
  large flows towards one end of the prominence or the other end. The siphon flow may go from the more magnetized end to the lower magnetized 
  end \citep{chen2014}.  In our case the bright arches could be considered as  a large structures containing siphon flows. \\

 How can we reconcile two magneto-hydrodynamic systems, one mainly horizontal and one turbulent? This  question was also put forward by \citet{Priest2014}. 
It should be important to work on the interpretation of the Stokes parameters profiles in this framework. The first test presented in this
work is encouraging;  while with the present data one cannot invert more complex magnetic models, we can at least propose models compatible with both the
polarimetric and the  imaging observations. \\

Finally, IRIS reveals the complexity of  MgII line profiles in an otherwise quiescent  prominence (the filament was quiescent with no network environment). Even 
when we consider
emission profiles having only one peak, it is difficult to interpret them
in terms of the existing models
\citep{Paletou1993} because these models show mostly reversed
profiles. Other observed profiles
exhibit a multipeak structure, which we interpret as due to the
line-of-sight Doppler shifts of individual
emission profiles. In a  future  paper we plan to perform a detailed
quantitative analysis of both types of
MgII profiles, using the existing non-LTE codes.
Moreover, the future modeling should consistently
explain the emission in the H$\alpha$ line
which we also present in this paper; H$\alpha$ can thus provide an
important constraint on the modeling of MgII lines.


\begin{acknowledgements}
IRIS is a NASA small explorer mission developed and operated by LMSAL with mission operations
executed at NASA Ames Research center
and major contributions to downlink communications
funded by the Norwegian Space Center
(NSC, Norway) through an ESA PRODEX
contract. H. T. is supported through contract 8100002705 from LMSAL to SAO.
Hinode is a Japanese mission developed and launched by ISAS/JAXA, with NAOJ as domestic partner and NASA and STFC (UK) as international partners. It is operated by these agencies in co-operation with ESA and NSC (Norway). SDO data are courtesy of NASA/SDO and the AIA science team. T.K. thanks NASA's LWS program for support. We thank also the team of THEMIS  and particularly B. Gelly, the director of THEMIS allowing us to obtain  coordinated observations of this prominence. We would like to thank  D. Shine for  providing  the  HINODE/SOT data, D.Crussaire, and R. Le Cocguen  for the observations in the solar tower in Meudon. 
We thank deeply Peter Martens and Petr Heinzel for their fruitful comments which help us to improve the manuscript. H.T. is supported by contract 8100002705 from LMSAL to SAO.
\end{acknowledgements}


\bibliographystyle{aa}    
\bibliography{references.bib}  

\begin{center}
 {\bf Appendix}
\end{center}
%
\begin{figure*} 
\centerline{\includegraphics[width=0.8\textwidth,clip=]{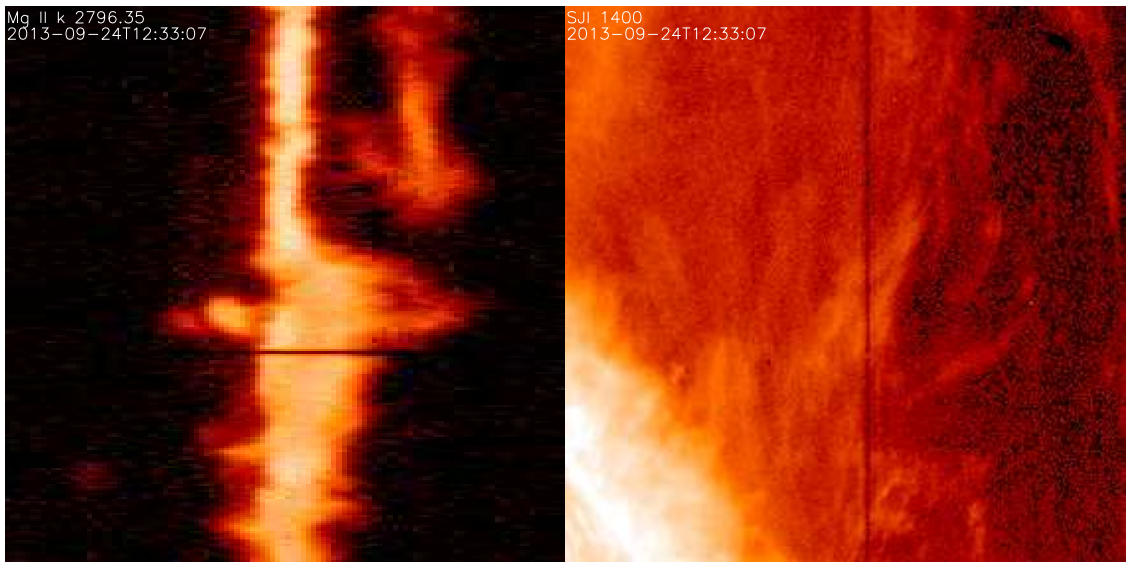}}
\caption{Figure online  explaining  the IRIS movies:   ({\it  left side})   Mg II k line spectra,   ({\it  right side})  IRIS Si IV slit-jaw image, the vertical line is the slit  position of the spectra. The wavelength range of the Mg II spectra is 2795.00 - 2797.78 \AA. The FOV of the  Si IV slit-jaw image  is  47"x47".  }
 \label{movie}
 \end{figure*}

\end{document}